\documentclass[%
 reprint,
 onecolumn,
superscriptaddress,
 amsmath,amssymb,
 aps,
floatfix
]{revtex4-2}

\pdfoutput=1
\usepackage{comment}
\usepackage{graphicx}
\usepackage{dcolumn}
\usepackage{bm}
\usepackage[version=4]{mhchem}
\usepackage{xcolor}
\usepackage{todonotes}
\usepackage{ulem} 
\usepackage{hyperref}
\usepackage{color,soul}
\usepackage{array} 
\hypersetup{hidelinks,backref=true,pagebackref=true,hyperindex=true,colorlinks=true,breaklinks=true,citecolor=blue,urlcolor= blue,linkcolor=cyan,bookmarks=true,bookmarksopen=false,pdftitle={`Title},pdfauthor={Author}} 

\begin{document}

\preprint{3D spin order in superconducting PrNiO$_2$}


\title{\textcolor{black}{Spin and orbital excitations in undoped infinite-layer superconducting PrNiO$_2$ and insulating CaCuO$_2$}}

\author{F.\,Rosa}
 \email[]{francesco1.rosa@polimi.it}
 \altaffiliation[Present address: ]{Institute of Physics II - University of Cologne, Zülpicher Straße 77, 50937 Köln, Germany}
 \affiliation{Dipartimento di Fisica, Politecnico di Milano, piazza Leonardo da Vinci 32, I-20133 Milano, Italy}

  \author{H.\,Sahib}
\altaffiliation[Present address: ]{Department of Physics, College of Science, University of Halabja, Halabja, Iraq}
\affiliation{Université de Strasbourg, CNRS, IPCMS UMR 7504, F-67034 Strasbourg, France}

\author{G.\,Merzoni}
 \affiliation{Dipartimento di Fisica, Politecnico di Milano, piazza Leonardo da Vinci 32, I-20133 Milano, Italy}
 \affiliation{European XFEL, Holzkoppel 4, Schenefeld, D-22869, Germany}

 \author{L.\,Martinelli}
\altaffiliation[Present address: ]{Physik-Institut, Universit\"{a}t Z\"{u}rich, Winterthurerstrasse 
190, CH-8057 Z\"{u}rich, Switzerland}
\affiliation{Dipartimento di Fisica, Politecnico di Milano, piazza Leonardo da Vinci 32, I-20133 Milano, Italy}
 
\author{R.\,Arpaia}
\affiliation{Department of Molecular Sciences and Nanosystems, Ca’ Foscari University of Venice, I-30172 Venice, Italy}
\affiliation{Quantum Device Physics Laboratory, Department of Microtechnology and Nanoscience, Chalmers University of Technology, SE-41296 Göteborg, Sweden}
 
 
\author{N.B.\,Brookes}
\affiliation{ ESRF, The European Synchrotron, 71 Avenue des Martyrs, CS 40220, F-38043 Grenoble, France}

\author{D.\,Di Castro}
\affiliation{Dipartimento di Ingegneria Civile e Ingegneria Informatica, Università di Roma Tor Vergata,
 Via del Politecnico 1, I-00133 Roma, Italy}
\affiliation{CNR-SPIN, Università di Roma Tor Vergata, Via del Politecnico 1, I-00133 Roma, Italy}

\author{K.\,Wohlfeld}
 \affiliation{Institute of Theoretical Physics, Faculty of Physics, University of Warsaw, Pasteura 5, PL-02093 Warsaw, Poland}

\author{M.\,Zinouyeva}
 \affiliation{Dipartimento di Fisica, Politecnico di Milano, piazza Leonardo da Vinci 32, I-20133 Milano, Italy}


\author{M.\,Salluzzo}
\affiliation{CNR-SPIN, Complesso Monte Sant’Angelo-Via Cinthia, I-80126 Napoli, Italy}

\author{D.\,Preziosi}
\affiliation{Université de Strasbourg, CNRS, IPCMS UMR 7504, F-67034 Strasbourg, France}

\author{G.\,Ghiringhelli}
\email[]{giacomo.ghiringhelli@polimi.it}
\affiliation{Dipartimento di Fisica, Politecnico di Milano, piazza Leonardo da Vinci 32, I-20133 Milano, Italy}
\affiliation{CNR-SPIN, Dipartimento di Fisica, Politecnico di Milano, I-20133 Milano, Italy}

\date{\today}

\begin{abstract}
Infinite-layer nickelates are among the most promising cuprate-akin superconductors, although relevant differences from copper oxides have been reported. Here, we present momentum- and polarization-resolved RIXS measurements on \textcolor{black}{nominally} undoped, superconducting PrNiO$_2$, and compare its magnetic and orbital excitations with those of the reference infinite layer cuprate CaCuO$_2$. In PrNiO$_2$, the in-plane magnetic exchange integrals are smaller than in CaCuO$_2$, whereas the out-of-plane values are similar, indicating that both materials support \textcolor{black}{three-dimensional antiferromagnetic correlations}. Orbital excitations, associated to the transitions within $3d$ states of the metal, are well reproduced within a single-ion model and display similar characteristics, except the Ni-$d_{xy}$ peak that, besides lying at significantly lower energy, shows an opposite dispersion to that of Cu-$d_{xy}$. \textcolor{black}{We attribute this difference} to the values of the orbital superexchange coupling between nearest neighbor sites. Our observations \textcolor{black}{suggest} that infinite-layer cuprates and nickelates share most of the spin and orbital properties, but in nickelates the larger charge-transfer energy $\Delta$ and smaller hopping integrals imply smaller spin-fluctuation energy and stronger localization of the doping charge on the metal site.

\begin{description}
\item[DOI]
https://doi.org/10.1038/s43246-026-01266-y
\end{description}
\end{abstract}

\maketitle

\section{Introduction}
The quest for high-$T_\text{c}$ superconductivity has brought to the attention of the condensed matter community the wide family of strongly correlated materials \cite{fradkin2015colloquium,keimer2015quantum,zaanen1990systematics}. Infinite-layer (IL) nickelates recently gained strong interest \cite{botana2022low,kitatani2020nickelate,botana2020similarities,nomura2022superconductivity, jiang2019electronic,goodge2021doping,zeng2020phase,hepting2020electronic} thanks to their apparent analogies with cuprates, including a square lattice-based structure dominated by superexchange antiferromagnetic interaction \cite{lu2021magnetic,ortiz2022magnetic} and the emergence of superconductivity upon hole doping.  
So far, the latter has been observed only in thin films \cite{li2019superconductivity,zeng2022superconductivity,chow2025bulk,osada2021nickelate,osada2020superconducting} which makes it difficult to assess the magnetic structure of the IL nickelate system, as no direct correspondence with the bulk can be established.  The bulk is mostly characterized by the absence of long-range order and the presence of a spin-glass behavior \cite{ortiz2022magnetic,zhou2025origin,dahabdahab2025unveiling}. Conversely, evidence of spin-spin correlations was reported from extensive Resonant Inelastic X-ray Scattering (RIXS) \cite{rossi2021orbital,lu2021magnetic}, as well as, X-ray Magnetic Circular Dichroism (XMCD) \cite{krieger2024signatures} experiments in superconducting and non superconducting thin films.


\begin{figure*}[]
\centering
\includegraphics[width=\textwidth]{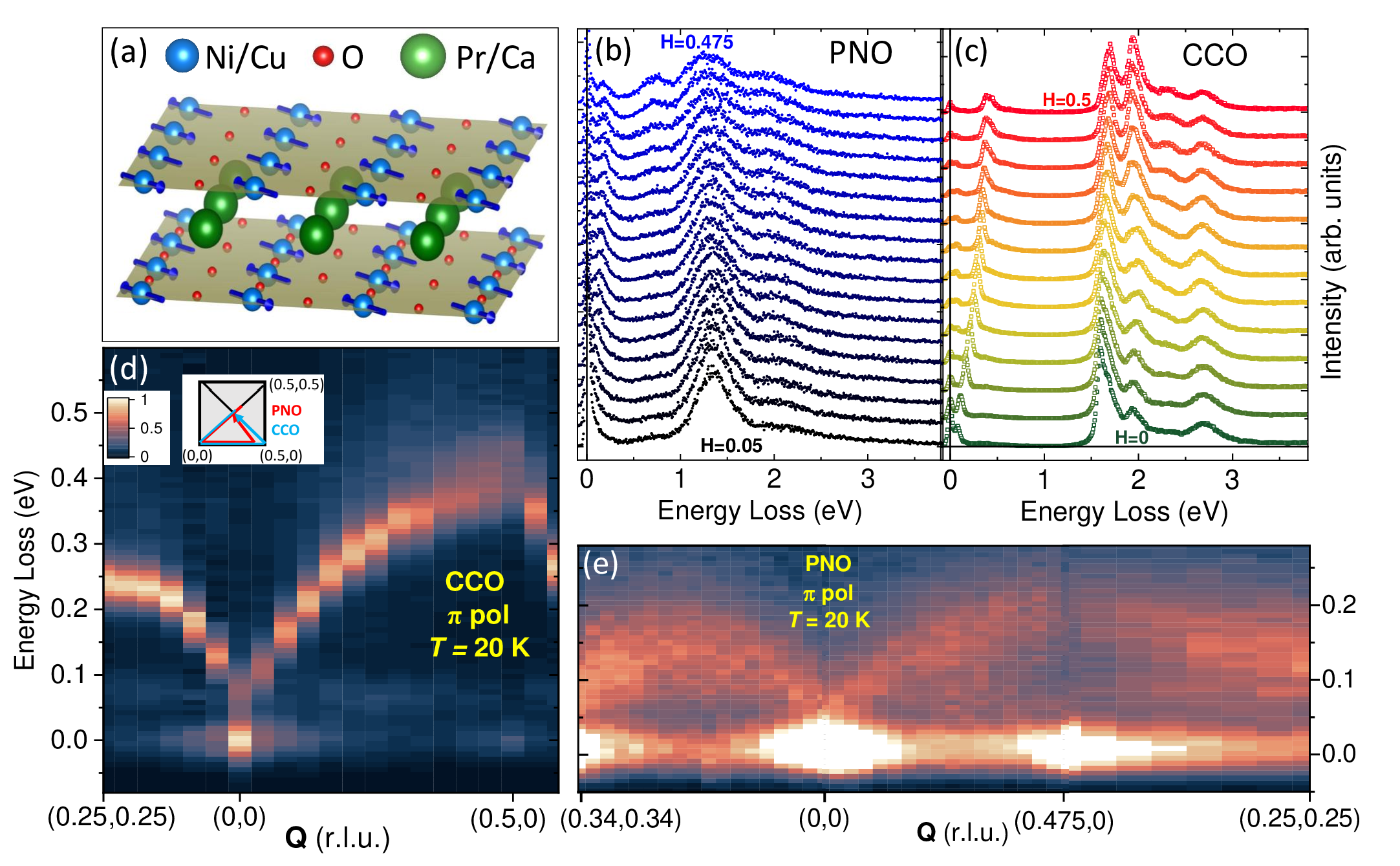}
\caption{\label{PNOCCOdata} \textbf{Comparison of RIXS spectra on PNO and CCO.} (a) Infinite-layer structure (ochre planes) of PNO \textcolor{black}{(CCO)}, showing the local tetragonal coordination of the nickel \textcolor{black}{(copper)} atoms to the surrounding oxygens; the dark blue arrows on metal sites depict antiferromagnetic spin order \textcolor{black}{(only reported for CCO)}. (b-c) RIXS spectra stacks for PNO and CCO respectively, along the $[H,0]$ cut of the Brillouin zone. (d-e) Momentum-dependent RIXS maps of CCO (d) and PNO (e) along the triple paths of the Brillouin zone shown in the (d) inset for the two samples. All (d-e) spectra have been normalized to the $dd$ energy integral along the interval $(1;3.5)$ eV.
} 
\end{figure*}

On the other hand, the analogies between cuprates and IL nickelates \cite{botana2020similarities,lee2004infinite} are not yet sufficient to prove the existence of a single underlying mechanism for unconventional superconductivity \cite{been2021electronic,worm2022correlations,nomura2022superconductivity}, but rather urge further investigation. At the root of the observed affinities is the common $3d^9$ electronic configuration of Cu$^{2+}$ and Ni$^{1+}$ ions occupying the corners of a planar spin 1/2 square lattice, with oxygen ligands along the edges. 

\textcolor{black}{
Within the Zaanen–Sawatzky–Allen framework \cite{zaanen1985band}, earlier works \cite{jiang2019electronic,hepting2020electronic,rossi2021orbital,botana2022low,chen2022electronic} tended to regard IL nickelates as genuine Mott-Hubbard insulators, where the ligand-to-metal charge-transfer energy $\Delta$ is larger than the Coulomb repulsion $U$, in contrast with charge-transfer insulators (like cuprates), where $\Delta<U$. A more recent interpretation, originating from RIXS studies on bi- and tri-layer nickelates \cite{lin2021strong,shen2022role,shen2023electronic} supports a different picture, revealing a mixed charge-transfer/Mott–Hubbard character that may extend to infinite-layer compounds, with $\Delta$ not much larger than $U$. This is consistent with the comparable magnitude of the magnetic exchange interaction in cuprates and nickelates, sensitively depending on $\Delta$ ($J\propto1/\Delta^3$ \cite{weihe2000superexchange}). In both scenarios, the oxygen $2p$ states lie at lower energy than in cuprates, and doped holes are expected to have a stronger metal character, although the precise degree of hybridization remains under debate. Moreover, the role of rare-earth $5d$ states has been proposed to generate Fermi-surface pockets \cite{been2021electronic,osada2021nickelate,kapeghian2020electronic,hepting2020electronic} and enable self-doping even in nominally undoped compounds \cite{jiang2019electronic,yang2022self,zhang2020self,lechermann2020late}, but recent angle-resolved photoemission measurements have not found clear evidence of such features at the $\Gamma$ point \cite{ding2024cuprate,li2025observation}.
}


As for magnetic properties, spin fluctuations are considered to play a major role in the formation of Cooper pairs in cuprates \cite{worm2024spin,scalapino2012common}, and a coexistence of magnetism and superconductivity has been suggested by XMCD \cite{krieger2024signatures}, as well as muon spin rotation/relaxation in cuprates and IL nickelates \cite{saykin2025spin,fowlie2022intrinsic}: therefore, a comparative investigation by other methods is timely.

Here, we used RIXS to investigate the dynamical spin response of the nominally undoped though superconducting IL nickelate PrNiO$_2$ (PNO) \cite{sahib2025superconductivity} and compare it to that of CaCuO$_2$ (CCO) \cite{stellino2025role,di2015high}, which has similar crystalline structure but fully insulating properties. \textcolor{black}{Despite} self-doping and superconductivity, in PNO the spin excitation peak is sharp and disperses similarly to CCO. We find that the in-plane exchange integrals are smaller in PNO than in CCO, consistently with previous reports \cite{lu2021magnetic}, while the out-of-plane values are comparable. Polarization-resolved RIXS reveals a better agreement of the PNO orbital excitations ($dd$s) with single-ion cross-section calculations, while in CCO a fourth peak occurs, whose origin remains unclear. Interestingly, the $d_{xy}$ peak displays an opposite dispersion in the two materials, which we interpret as a consequence of distinct orbital superexchange couplings driving the orbiton propagation in the two materials: nearest neighbor for the nickelate, next-nearest neighbor for the cuprate \cite{martinelli2024collective}.


\begin{figure*}[]
\centering
\includegraphics[width=1\textwidth]{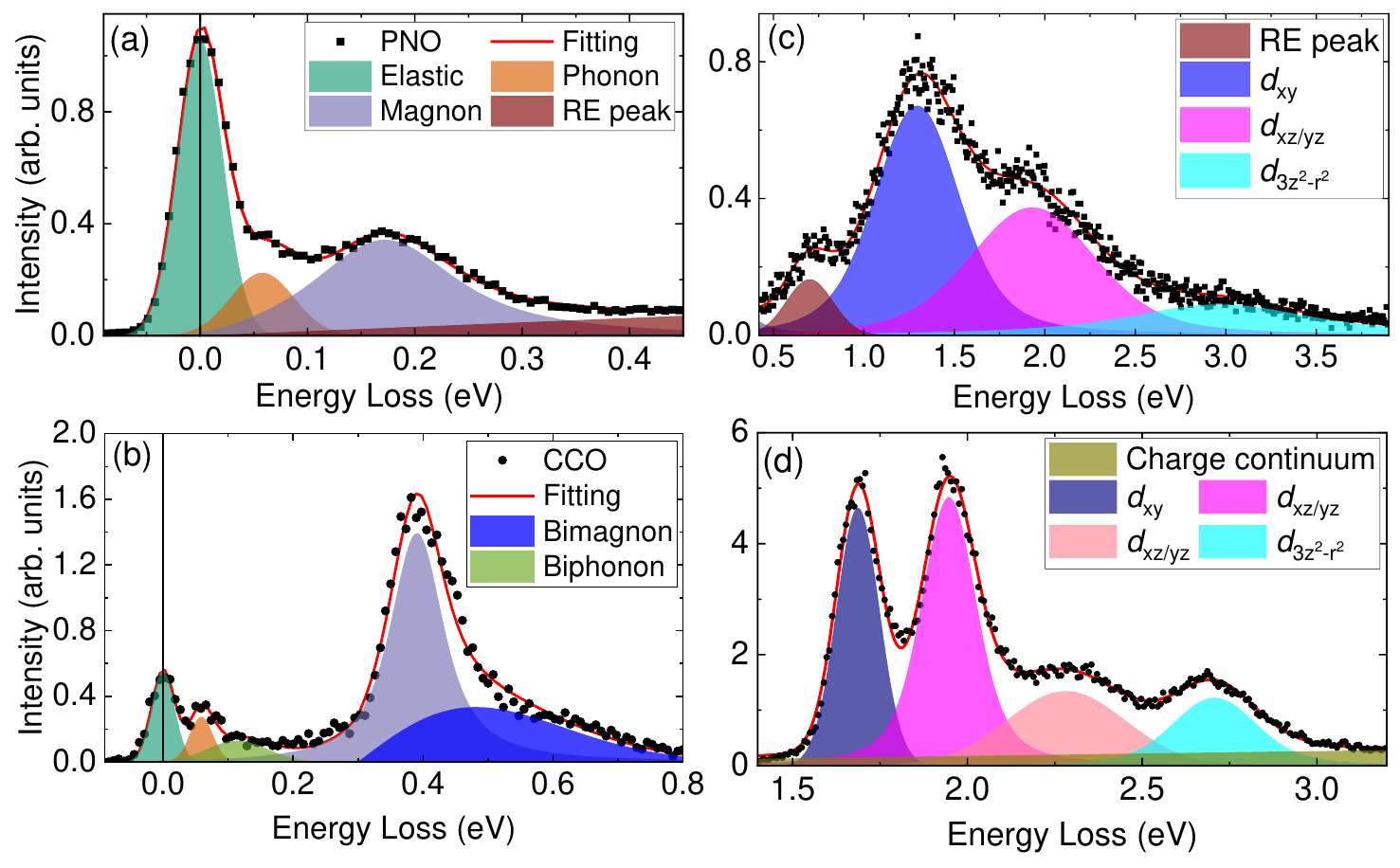}
\caption{\label{fitComparison} \textbf{Analysis of RIXS spectra.} (a) Fitting of PNO spectrum with $\pi$ incident polarization, as described in the main text. Data points were obtained as the sum of different spectra along $[H,0]$ with $H$ between 0.4 and 0.475 r.l.u. at incident energy 852.48 eV. (b) Same fit for CCO, with $H=0.465$ r.l.u.; a second Gaussian is added here (light green) representing the phonon overtone, while the rare-earth peak tail has been replaced by the bimagnon continuum (dark blue). Elastic, phonon and magnon colors are the same as in (a). (c-d) Same fittings as (a-b) respectively, but focusing on the orbital excitations area.
} 
\end{figure*}

\section{Results and discussion}
Figure \ref{PNOCCOdata} displays an overview of RIXS spectra for both samples, showcasing orbital and spin excitations. 
\textcolor{black}{The common infinite-layer structure is depicted in Panel (a), which can represent either CCO or PNO depending on the assignment of Cu/Ni and Ca/Pr atoms. The antiferromagnetic spin lattice, also shown, has so far been reported only for the cuprate.}
Panels (b-c) report spectra taken at different transferred momenta along the [$H$,0] cut of the 2D reciprocal space. The orbital ($dd$) excitations at high energy range (1-3 eV) show a small but significant dispersion as a function of the in-plane momentum, as discussed below \cite{sala2011energy,martinelli2024collective}. 
\textcolor{black}{For the nickelate, an additional feature is observed at 0.7 eV. This excitation has already been reported in the literature \cite{rossi2021orbital,gao2024magnetic, hayashida2024investigation} and becomes progressively enhanced towards the Brillouin zone edge. Its out-of-plane character is further supported by the fact that it is more enhanced for $\pi$-polarized incident light than $\sigma$ (see also Figure \ref{polarimeter} below). This feature has mainly been interpreted \cite{hepting2020electronic} in terms of hybridization between Ni $3d$ and rare-earth $5d$ states. Here we adopt this interpretation, although alternative scenarios cannot be ruled out. Indeed, a possible extrinsic origin is supported by the fact that the peak appears stronger in samples showing a more intense diffraction signal from impurity phases \cite{tam2022charge}.}
The sharp feature dispersing up to $\sim 200$ meV in PNO and $\sim 400$ meV in CCO is assigned to $\Delta S=1$ spin excitations, in agreement with previous RIXS work \cite{lu2021magnetic,krieger2022charge,rosa2024spin,rossi2024universal,peng2017influence,peng2018dispersion,martinelli2022fractional}. These spin excitations follow a single-magnon dispersion across most of the reciprocal space along the M-$\Gamma$-X path (panels d-e). Whereas the sharpness of the magnetic peak in CCO is a direct consequence of the long-range antiferromagnetic (AFM) order of the undoped compound  \cite{braicovich2009dispersion,martinelli2022fractional}, the coexistence of superconductivity and sharp spin excitations in IL nickelates demonstrates the mild effect of self-doping on magnetism. 
This constitutes an important difference with respect to cuprates, where the chemical hole doping, required for superconductivity, quickly disrupts spin order and progressively smears the magnon excitation peak \cite{peng2018dispersion} across the superconducting dome (see below for a more quantitative comparison). Our observations are consistent with the muon spin rotation measurements in superconducting IL nickelates by Fowlie et al. \cite{fowlie2022intrinsic}, which hint at a similar coexistence. A recent debate was sparked by the observation of an elastic intensity increase around the quasi-commensurate wavevector $Q\sim(0.33,0)$. This was first interpreted as an evidence of charge order analogous to the Charge Density Waves observed in cuprates \cite{rossi2022broken,krieger2022charge,tam2022charge}, but later dismissed as resulting from periodic arrangement of residual apical oxygens from incomplete topotactic reduction \cite{raji2023charge,parzyck2024absence}. Our integrals of the spectral weight in the quasielastic region, displayed in Figure S1 of the Supplementary Material, show no evidence of such an increase, confirming the absence of charge order in our samples and the fully occurred topotactic reaction.

\begin{figure*}[]
\centering
\includegraphics[width=1\textwidth]{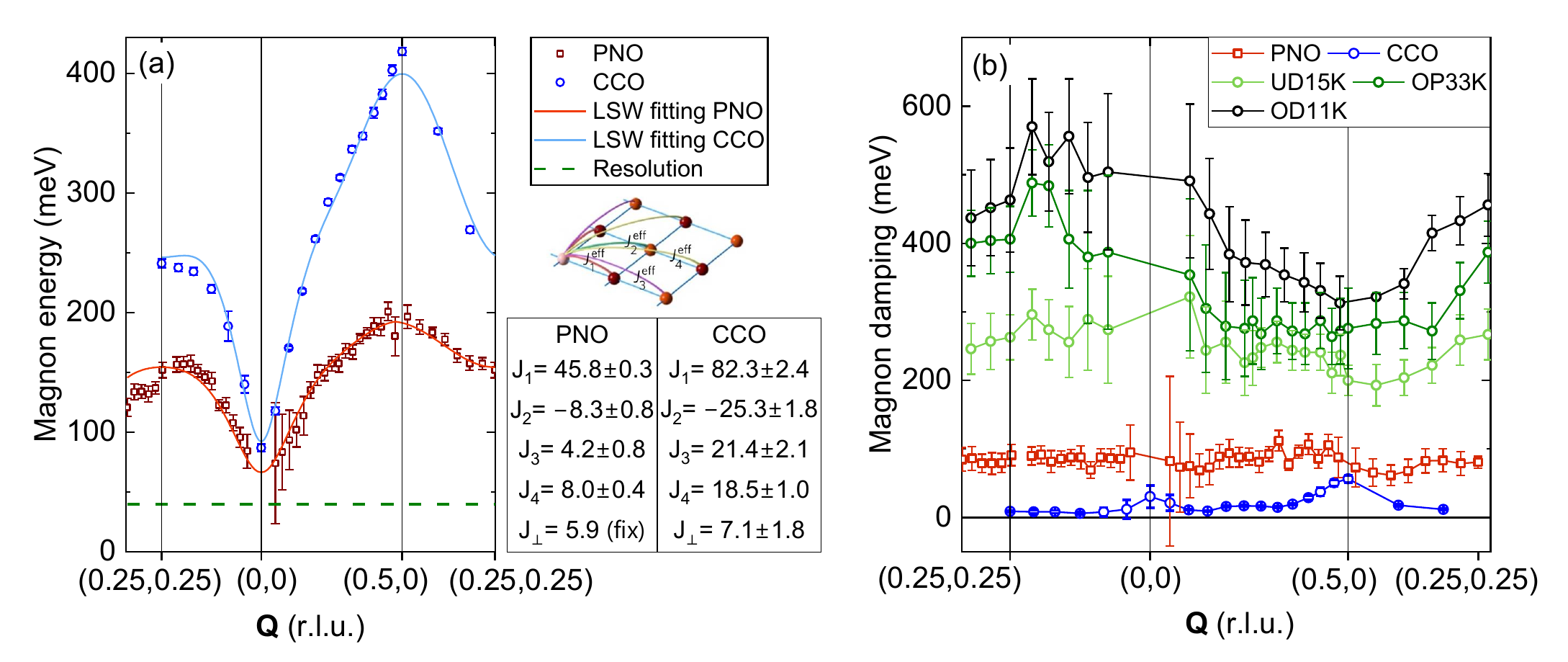}
\caption{\label{LSWfits} \textbf{Comparison of magnon energy and damping.} (a) Fitting of dispersions of PNO and CCO with the Linear Spin Wave model fit. The definitions and values (in meV) of the four in-plane exchange coupling constants are displayed on the right \cite{peng2017influence}. (b) Comparison of the damping coefficients of PNO and CCO with those extracted from analogous fittings on Bi2201 cuprate at different doping levels in Ref. \onlinecite{peng2018dispersion} (UD = underdoped, OP=optimally doped, OD=overdoped). Error bars in both panels represent 95\% confidence intervals from the fittings.} 
\end{figure*}

In order to determine the energy of the magnetic excitations, we fitted the elastic and phonon peaks with resolution-limited Gaussian functions, and the magnetic peak with the damped harmonic oscillator spectral shape, convolved with a Gaussian accounting for the experimental resolution \cite{peng2018dispersion} (see Figure \ref{fitComparison}(a-b)). A phonon peak at $\approx 80$ meV \textcolor{black}{was included in the spectral fitting.} Phonon and magnon overtones \textcolor{black}{were considered only} in the cuprate fit, since in the nickelate these features are too broad and too close in energy \textcolor{black}{to be reliably disentangled}. 
A similar fit was performed on the high-energy peaks related to the orbital transitions between $3d$ levels (Panels (c-d), to be discussed in the following). More details about the fittings are available in the Supplementary Information \ref{SupplementMater}). The momentum-energy dispersions for PNO and CCO are reported in Figure \ref{LSWfits}(a), while \ref{LSWfits}(b) shows a comparison between the extracted damping coefficients, related to the spectral broadening of the magnon, and the corresponding values for superconducting (Bi,Pb)$_2$(Sr,La)$_2$CuO$_{6+\delta}$ (Bi2201) samples at different doping levels, taken from Ref. \onlinecite{peng2018dispersion}. The peak broadening for the nickelate is found to be nearly constant ($\sim100$ meV) over the explored momentum range. Such value is obviously larger than for CCO, where \textcolor{black}{the magnetic peak is resolution-limited: in this case} long-range AFM order provides almost undamped spin wave propagation. At the same time, it is less than half the corresponding value for the lowest doping level of superconducting Bi2201 ($\gtrsim200$ meV, within the confidence intervals). This demonstrates that self-doping has a much milder impact on spin order than chemical doping, endowing IL nickelates with a non-disruptive way to achieve superconductivity which is totally absent in copper oxides.
\\We notice also an increase of the elastic intensity around the X point, a result previously reported but not discussed, for PNO and NdNiO$_2$  \cite{lu2021magnetic,rossi2024universal}. 
Upon approaching the X point in (0.5,0), the spin excitation becomes less intense and its shape is not an individual peak directly identified with a single magnon, as a consequence of fractionalization phenomena, related to the square lattice and independent of the material \cite{dalla2015fractional}, and of multi-magnon excitations \cite{betto2021multiple}. At the same time a continuum, possibly due to spinon pairs, appears and becomes increasingly intense towards the X point: for CCO these effects are very evident and were previously discussed by Martinelli et al. \cite{martinelli2022fractional}. For PNO and other IL nickelates the broadening is harder to determine but the loss of intensity was previously observed \cite{rossi2024universal}. We consider that here a single-peak fit is accurate enough for our purpose, although for CCO it results in an overestimation of the single-magnon damping energy close to (0.5,0) as shown in Figure \ref{LSWfits}(a).

\begin{figure*}[]
\centering
\includegraphics[width=\textwidth]{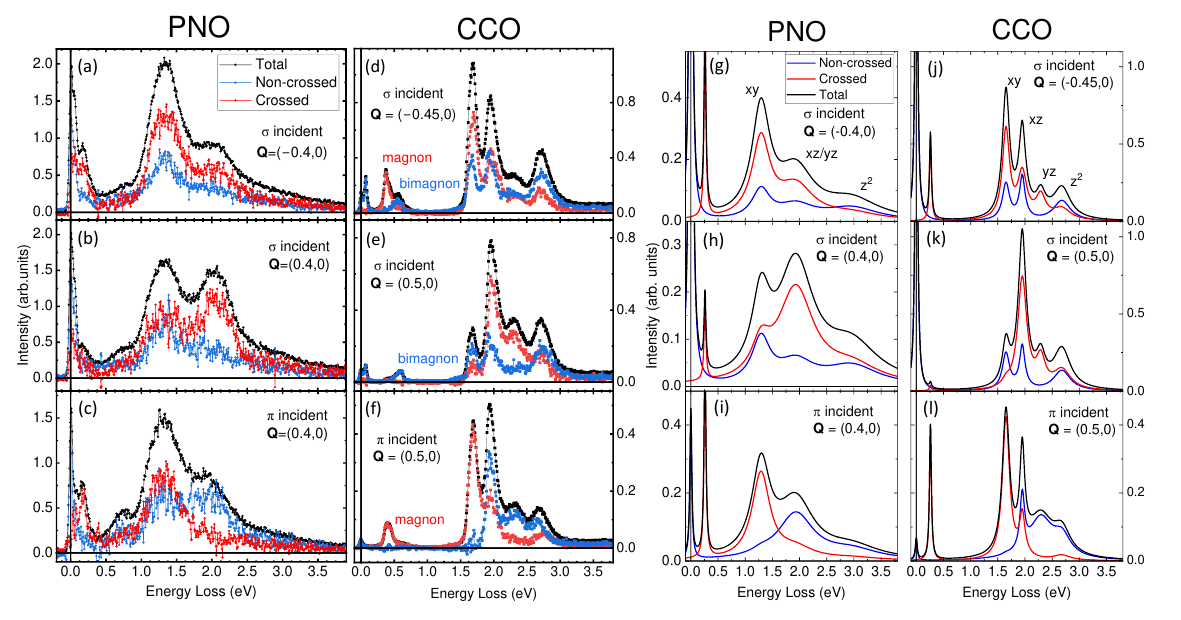}
\caption{\label{polarimeter} \textbf{Comparison between polarimetric RIXS data and theory.} (a-f) Polarization-resolved RIXS spectra for PNO (a–c) and CCO (d–f), measured under the geometry and polarization conditions specified into each panel. (g–l) Corresponding single-ion RIXS cross-section calculations for the same polarization and geometry conditions as in (a–f), respectively.
} 
\end{figure*}

To analyze the differences between the two samples, we fitted the extracted dispersion curves with those predicted by Linear Spin Wave theory (LSW) \cite{coldea2001spin}, \textcolor{black}{as implemented in} the SpinW library in Matlab \cite{spinWsite}. We \textcolor{black}{included} four in-plane exchange couplings $J_{1-4}$ and one out-of-plane \textcolor{black}{term} $J_{\perp}$, \textcolor{black}{following the strategy previously adopted for CCO in Ref.} \onlinecite{peng2017influence}. This is \textcolor{black}{new} for IL nickelates, whose magnon dispersion \textcolor{black}{has so far been modeled using purely two-dimensional LSW approaches}  \cite{lu2021magnetic,gao2024magnetic,rossi2024universal,yan2025persistent} or, at most, \textcolor{black}{by introducing} an out-of-plane magnetic anisotropy  \cite{bialo2024strain} which is however difficult to \textcolor{black}{constrain} from RIXS spectra. \textcolor{black}{As for CCO, exchange couplings up to fourth nearest neighbors were needed to reproduce the zone-boundary energy.} The inclusion of $J_{\perp}$ is motivated by the fact that, upon approaching $\Gamma$ at non-integer $L$, the single-magnon energy remains large, closely resembling the behavior found in CCO. \textcolor{black}{For PNO, the $J_{\perp}$ value was not directly estimated as a fitting parameter but was kept fixed during the procedure. The reason lies in the difficulty to safely estimate the value of the gap at the $\Gamma$ point (see Supplementary Material for more details, in particular Figures \ref{lowHfittings} and \ref{JperpPNO_fig}). The value 5.9 meV reported in the figure provided the best fitting quality. Nevertheless, we can reasonably assume $J_{\perp}$ for PNO to be between 0 and 10 meV.} The values of the extracted exchange $J$s (in meV) are reported in Figure \ref{LSWfits}(a). The nearest-neighbor exchange $J_1$ is about twice as large in CCO (82 meV) than in PNO (46 meV), in agreement with the literature \cite{been2021electronic,lu2021magnetic,rosa2024spin}. Conversely, $J_{\perp}$ is comparable in the two compounds (7 meV for CCO and 6 meV for PNO), leading to out-to-in-plane exchange ratio $J_{\perp}/J_1$ of $8.5\%$ for CCO and $13\%$ for PNO. We notice that these values of $J_{\perp}$ are close to the intra-unit-cell coupling of bi-layer cuprates like YBCO \cite{peng2017influence}. The absence of apical oxygens in IL nickelates leads to \textcolor{black}{a} long-range, in-plane hopping \textcolor{black}{phenomenon qualitatively} similar to cuprates', despite the larger $\Delta$. We observe that the 3D character of the spin-spin correlation is also similar in the IL nickelates and in the CCO, as testified by the comparable value of the spin excitation energy at in-plane zone center and non-integer $L$ (60-80 meV respectively). \textcolor{black}{Whether rare-earth ions play a substantial role in this mechanism, as proposed in several (mainly theoretical) studies \cite{peng2023charge,zhang2023rare,been2021electronic}, remains unclear, since other works suggest a much weaker influence \cite{osada2021nickelate}.} \textcolor{black}{Moreover, our results} give a proof of the fully occurred topotactic reduction of the PNO samples and the absence of interstitial oxygen atoms.


\textcolor{black}{A possible role of epitaxial strain can also be considered, since the two films experience opposite strain conditions: PNO is grown on SrTiO$_3$ ($a_{\mathrm{STO}}=3.905$\,\AA), whose lattice parameter is smaller than that of PNO ($a_{\mathrm{PNO}}=3.94$\,\AA), whereas CCO is grown on NdGaO$_3$ ($a_{\mathrm{NGO}}=3.861$\,\AA), whose lattice parameter is only slightly larger than that of CCO ($a_{\mathrm{CCO}}=3.856$\,\AA). This results in a compressive strain of about +0.9\% for PNO and a very small tensile strain of about -0.1\% for CCO. Epitaxial strain has been shown to affect the in-plane magnetic exchange in both cuprates \cite{minola2013measurement,ivashko2019strain} and non-infinite-layer nickelates \cite{bialo2024strain,chen2024electronic,zhong2025epitaxial}, with compressive (tensile) strain generally reducing (enhancing) the exchange interaction. In our samples, however, strain would tend to increase $J_1$ in PNO and slightly reduce it in CCO, thereby opposing the observed difference rather than causing it. Since the strain mismatch is also too small to account for the sizable difference in the extracted exchange values, the smaller $J_1$ of PNO relative to CCO must be intrinsic.}

Along the $[H,H]$ direction, we notice that the maximum of the dispersion in PNO is not at the BZ boundary (0.25,0.25) r.l.u., but rather at (0.2,0.2) r.l.u., and that the elastic peak intensity increases very much beyond the BZ boundary in (0.25,0.25) r.l.u. (see Figure \ref{PNOCCOdata}(e)). These facts might suggest the presence of an incommensurate resonant diffraction reflection distinct from the AFM (0.5,0.5) point, which cannot be reached in RIXS due to kinematic limitations. Those peculiarities had already been observed but not discussed in Ref. \onlinecite{rossi2024universal} for PNO and LaNiO$_2$.

To gain further insights, we measured RIXS spectra with analysis of the linear polarization of the scattered photons. We chose large momenta in the $[H,0]$ direction, with both positive and negative values of $H=\pm0.4$ r.l.u. for PNO and $H=\pm0.5$ r.l.u. for CCO. For positive $H$ values we took both $\sigma$ and $\pi$ incident polarization, for negative $H$ only the $\sigma$ polarization was used because cross-sections for $\pi$ incidence are much smaller \cite{sala2011energy}. The results are shown in Figure \ref{polarimeter}, where the spectral components with scattered polarization orthogonal to the incident one ($\sigma \pi'$ and $\pi \sigma'$, i.e., crossed polarization) are given in red and those with parallel polarization ($\sigma \sigma'$ and $\pi \pi'$) are in blue. In Figure \ref{polarimeter}(c)-(f), we see that the single magnon peak at positive $H$ has crossed character for both samples, in agreement with literature on cuprates \cite{fumagalli2019polarization,martinelli2022fractional} and with theory \cite{ament2009theoretical,sala2011energy}. With $\sigma$-polarized incident light and positive $H$, the magnetic peak in CCO is purely parallel, consistently with the transfer of an even number of angular-momentum units (bimagnon). In contrast, in the nickelate this feature — if present — is too weak and broad to be clearly distinguished (Panels (e) and (b), respectively). In all PNO spectra we see that an unpolarized continuum is present, likely resulting from self-doping, which leads to a not purely crossed polarization character of the magnon in panel (c) \textcolor{black}{and non purely parallel polarization} of the phonon peak in all spectra.

The differences between the two samples are more evident in the $dd$s spectral region (1-3 eV). Here the peaks have been fitted as already shown in Figure \ref{fitComparison}(c-d), and assigned according to the known crystal field splitting of $3d$ orbitals in a square tetragonal environment \cite{sala2011energy}. Disregarding the peak at 0.7 eV in PNO, the lowest energy feature can be assigned to the $d_{xy}$ orbital excitation in both samples. The energy position obtained by a multi-peak fitting is 1.29 eV and 1.65 eV for PNO and CCO respectively, with the difference likely being due to the different hopping parameters and in-plane lattice constants of the two materials, which are also grown on different substrates. 
The $d_{xz/yz}$ excitation in the cuprate appears split into two components; again this phenomenon has been already observed \cite{martinelli2024collective} but not explained in CCO, although it might not be due to simple splitting between the single $d_{xz}$ and $d_{yz}$ features (see also the comparison with cross-sections explained below). Nevertheless, the $d_{xz/yz}$ energy position is quite similar between the two systems (1.93 eV for PNO, 2.05 eV for the center of mass of the two features of CCO). \textcolor{black}{The $d_{z^2}$ feature in the nickelate is very broad and weak, so that its position is quite hard to estimate precisely. Although determining its exact position is not crucial for our purposes, fitting results suggest it is found at slightly larger energy than in the cuprate.} Polarization analysis shows that the $d_{xz/yz}$ features have similar character for PNO and CCO in all three geometries, while the $d_{xy}$ peak is similar at negative $H$ but is very different at positive $H$. In fact, for $H>0$ it has \textcolor{black}{predominant} crossed character for $\pi$ and parallel character for $\sigma$ for CCO but it is of mixed polarization character for PNO\textcolor{black}{, at odds with the completely crossed polarization expected from theory (see Figure \ref{polarimeter}(i)).} A possible intuitive explanation is the proximity of the 0.7 eV peak, which is fully unpolarized in all cases. \textcolor{black}{The fact that its influence appears maximum in the case of grazing emission and $\pi$ incidence, where we report the discrepancy, appears consistent with the out-of-plane nature of the peak, already observed in previous studies \cite{krieger2022charge}.} \textcolor{black}{Accepting the interpretation of the latter in terms of} Ni $3d$-rare earth $5d$ hybridized states \cite{hepting2020electronic}, the mixed polarization character of the 1.29 eV peak might be a hint of stronger hybridization of Pr $5d$ with Ni $3d_{xy}$ states  than with the $3d_{xz/yz}$ ones. However, recent photoemission measurements \cite{li2025observation} have shown no evidence of rare-earth states close to the Fermi level, highlighting instead a predominating role of electride-like interstitial $s$ states, where electrons are delocalized over several voids and coexist with the Ni $3d_{x^2-y^2}$ electrons. It is not excluded that such $s$ states can also interact with the $3d_{xy}$ states, which are oriented in the same plane as the $3d_{x^2-y^2}$: further investigation is required to verify this hypothesis.
\\By comparing polarimetric RIXS data with single-ion cross-sections calculations shown in panels (g-l), and based on the energies listed in Table \ref{tab:ddenergies}, almost all of the CCO spectra can be quite well reproduced. We added here a splitting of the $d_{xz/yz}$ peak into the two separate $d_{xz}$ and $d_{yz}$ features, giving good agreement except for the grazing incidence case. On the other hand, for the PNO spectra the measured $d_{xy}$ peak at $H=0.4$ for $\sigma$-polarized incident light appears significantly stronger than predicted by the calculations, which may indicate contributions from Pr $5d$ and/or interstitial $s$ states.

\begin{table}
    \begin{tabular}{m{2.7cm} m{2.7cm} m{2.7cm}}
     \hline
     \hline
      Orbital peak (eV) &   PNO    & CCO \\
      \hline
       $d_{xy}$      &   $1.29\pm0.02$   & $1.65\pm0.02$  \\
       $d_{xz/yz}$      &    $1.93\pm0.06$    & $1.95\pm0.01$  \\
                        &                   &   $2.29\pm0.06$ \\
       $d_{z^2}$      &    $2.96\pm0.58$    & $2.68\pm0.04$ \\
       \hline
       \hline
    \end{tabular}
    \caption{\textbf{Orbital peaks energies (in eV) for PNO and CCO.} Energy losses of the $dd$ peaks at BZ edge as extracted from the fittings in Figure \ref{fitComparison}(c-d).}
    \label{tab:ddenergies}
\end{table}

\begin{figure*}[]
\centering
\includegraphics[width=1\textwidth]{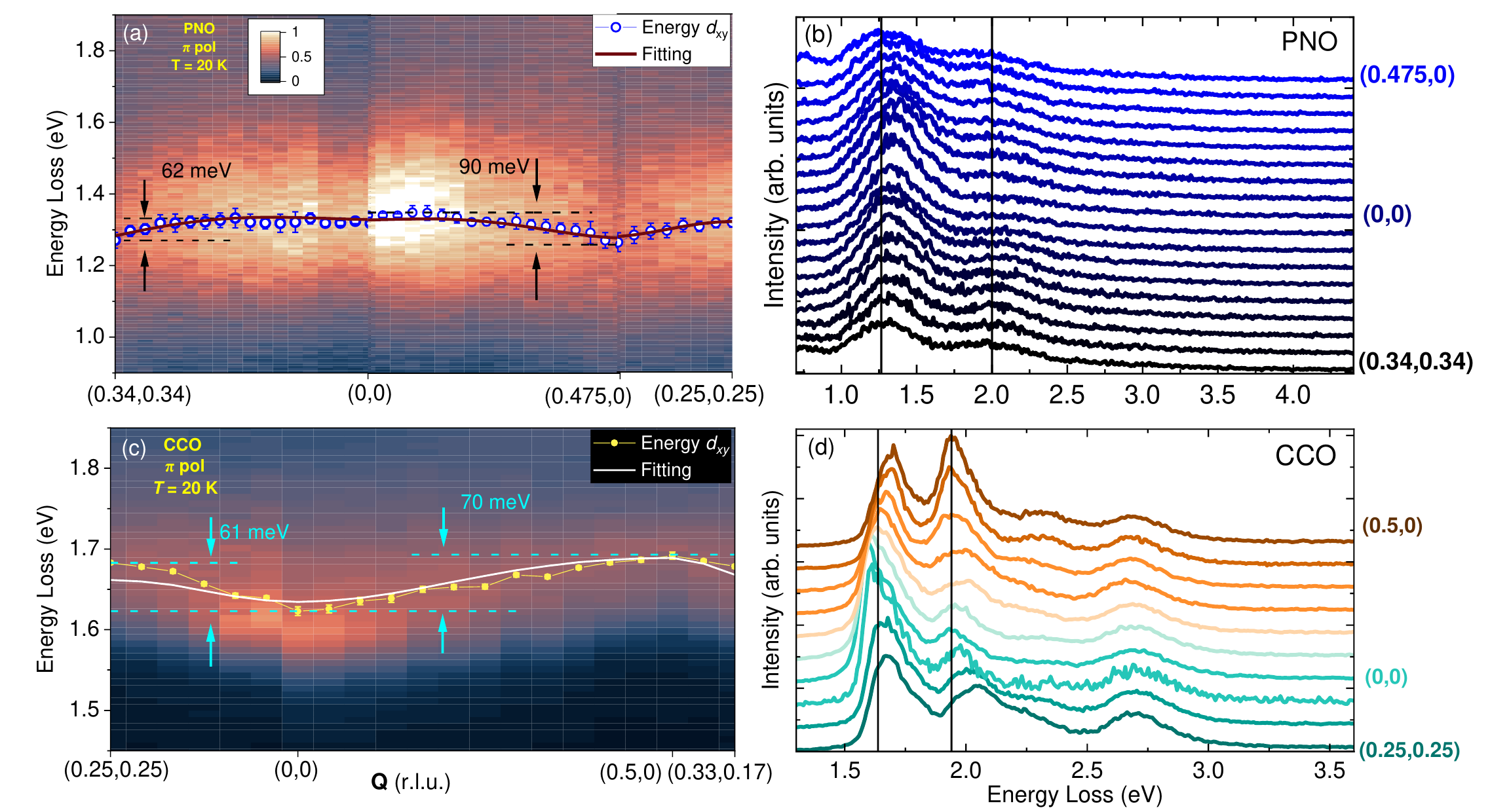}
\caption{\label{orbitonsFig} \textbf{Orbital excitation dispersion.} (a) Momentum-resolved map in the $d_{xy}$ energy range of PNO; dots highlight the dispersion of the peak, while the continuous line represents the charge transfer model fitting (see text). Horizontal dashed lines quote the maximum and minimum energies of each branch, with their distance being reported. The followed BZ cuts are the same shown in the inset of Figure 1(d). (b) Stack of spectra of PNO in the whole $dd$ range, with vertical lines fixed at the minimum of the dispersion of the $d_{xy}$ and $d_{xz/yz}$ peak. (c-d) Same as (a-b), for CCO. Error bars in (a)-(c) represent 95\% confidence intervals from the fittings.
} 
\end{figure*}

We conclude the comparison between the orbital excitations by examining their dispersion. If they had a perfectly local nature, they would show no momentum dependence, and in most layered cuprates, this is indeed what has been observed~\cite{sala2011energy,igarashi2012magnetic,fumagalli2019polarization}. For 1D cuprates it was theoretically proposed and experimentally \textcolor{black}{measured} \cite{schlappa2012spin,fumagalli2020mobile} that some dispersion can be attributed to an effective orbital exchange interaction (Kugel-Khomskii spin-orbital model)~\cite{kugel1982jahn} ultimately ascribed to the Cu-O hopping and to 1D fractionalization, which allows orbitons to propagate independently from magnetic excitations~\cite{wohlfeld2011intrinsic}. 
In 2D cuprates it has been shown theoretically that the mechanism is strongly frustrated by the antiferromagnetic spin order that would immediately dress the propagating orbital excitation with a trail of spin excitations~\cite{wohlfeld2012dispersion, wohlfeld2011intrinsic}. Such coupling between the magnetic and orbital degrees of freedom, which strongly hinders orbiton propagation by nearest neighbor (NN) orbital superexchange, is known as magnetic string effect and is well-known from the propagation of holes in antiferromagnets~\cite{Kane1989, Martinez1991}. However, in CCO the dispersion of the $d_{xy}$ state was recently observed: it is attributed to the large next-nearest neighbor (NNN) orbital superexchange, which allows the orbiton to propagate along the diagonal avoiding coupling to magnons~\cite{martinelli2024collective}. 
\textcolor{black}{In Fig.~\ref{orbitonsFig}(a,b) we show that in PNO the $d_{xy}$ peak disperses as a function of momentum. In comparison with the CCO case (c,d), the dispersion has similar bandwidth but different shape: the energy is maximal (minimal) at $\Gamma$ and decreases (increases) with momentum along both the $[H,0]$ and $[H,H]$ directions for PNO (CCO). To further elucidate the processes underlying the observed orbiton motion, we fit the experimentally observed $d_{xy}$ orbital dispersion in the two materials: the obtained fittings are shown as solid lines in Fig.~\ref{orbitonsFig}(a)–(c). For PNO, we use the following {\it a priori} rather generic and unbiased dispersion relation:
%
%
\begin{equation}
\label{orbitonDispersionNew}
\varepsilon_{\boldsymbol{k}} = \varepsilon_{\boldsymbol{0}} + 2t^{orb}_{1}[cos (2\pi k_x) + cos (2\pi ky)] 
+ 2 t^{orb}_{2} [cos (2\pi k_x)cos (2\pi ky)]  + 2 t^{orb}_3 [cos (4\pi k_x) + cos (4\pi k_y)],
\end{equation}
where $t^{orb}_{1}$ is the nearest-neighbor (NN) orbiton hopping, $t^{orb}_{2}$ is the next-nearest-neighbor (NNN) orbiton hopping, and $t^{orb}_{3}$ is the third-neighbor orbiton hopping. Then, the fitted values of the effective orbiton hoppings are $t^{orb}_{1} = 12.5\pm3.8$ meV, $t^{orb}_{2} = 0.2\pm4.8$ meV, $t^{orb}_{3} = 5.4\pm1.7$ meV, indicating that the dominant orbiton hopping occurs between nearest neighbors.
\\On the other hand, the orbital dispersion for CCO has already been observed and thoroughly explained \cite{martinelli2024collective} by proposing a dominant role of the NNN orbital exchange, following the strong covalence in CCO and suppression of the NN exchange caused by coupling to magnons. Here we follow this interpretation by fitting the CCO orbiton dispersion with the relation
\begin{equation}
\label{orbitonDispersionOld}
\varepsilon_{\boldsymbol{k}}=2t^{orb}_2cos(2\pi k_x)cos(2\pi k_y)
\end{equation}
retrieving for CCO a next-nearest neighbor hopping $t^{orb}_{2} = -13.6\pm3.5$ meV.
\\Before attempting to understand the obtained results using a microscopic model, let us stress the strong difference between the observed orbiton dispersion in the nickelate and in the cuprate. Whereas in the latter case the (negative) dominant orbital exchange occurs on the same antiferromagnetic sublattice, in PNO the nearest-neighbor hopping is both positive and dominant. This suggests that the origin of the orbiton dispersion in CCO and PNO must be quite different, as explained in detail in the two steps below.
\\\textit{First}, the much weaker covalence in PNO than in CCO ~\cite{botana2020similarities} implies that longer-range orbital superexchange interactions are significantly reduced. Therefore we expect that, unlike in CCO, the nearest-neighbor orbital superexchange in PNO strongly dominates over the much weaker next-nearest-neighbor contribution.
\\To validate this hypothesis, we estimated the NN and NNN orbital superexchange couplings, $J^{\mathrm{orb}}$, in PNO using a simple Hubbard-based three-band model, similar to that previously employed to estimate the orbital superexchange interactions in CCO~\cite{martinelli2024collective}. Details of this model can be found in the Supplementary Information of Ref.~\onlinecite{martinelli2024collective}. The corresponding model parameters are listed in Table~\ref{tab:modelParameters}, where the values for CCO are included for comparison. 
\\The choice of model parameters is crucial in this context, as we discuss them in some detail in the Methods section. In the end, by plugging such values from the three-band model into the orbital superexchange formulas given in Ref.~\cite{martinelli2024collective}, we find that the NN orbital superexchange is dominant in PNO, while the NNN exchange is relatively small and can be safely neglected below; see Table~\ref{tab:orbitalExchangeJs}. This is in stark contrast to CCO, where the NNN superexchange was found to be very large and was therefore the only exchange interaction included in the theoretical description~\cite{martinelli2024collective}. The origin of this difference between the two materials lies in the substantially stronger covalence in the cuprates as compared to the nickelates~\cite{botana2020similarities}.
\\\textit{Second}, having estimated the values of the orbital superexchange couplings in PNO, we are now left with the task of relating these results to the observed orbiton dispersion relations. Here, the first point to stress is that a large nearest-neighbor orbital superexchange drives nearest-neighbor orbiton motion, which is, as discussed above, strongly frustrated in antiferromagnets due to the spin-string effect. 
In other words, a large nearest-neighbor orbital exchange $J^{\mathrm{orb}}_1$ does not automatically imply a large free-orbiton hopping amplitude $t^{\mathrm{orb}}_1$. Indeed, the latter can be strongly suppressed by the string effect, i.e., by the strong scattering of the orbiton off magnons generated during its motion through the antiferromagnetic background.
\\Nevertheless, it turns out that there exists a `way out' and the effective orbiton hopping is actually quite large in PNO. The key point is that a finite Hund’s exchange in the intermediate state of the orbital exchange process allows for a concomitant orbital and spin flip, which leads to a finite orbiton hopping that preserves the antiferromagnetic order.
While one might expect such processes to be strongly suppressed, a relatively large ratio of the Hund's exchange $J_H$ to the Hubbard $U$ in PNO makes makes these processes remarkably efficient. Altogether, this leads to $t^{orb}_{1} \approx 0.6 J^{orb}_{1}$ in PNO; see the SM section for more details. Note that the orbiton motion is quite different in CCO~\cite{martinelli2024collective}.
Here the orbiton can move efficiently via a large $t_{2}^{orb} \approx J_{2}^{orb}$, as shown in~\cite{martinelli2024collective}. 
\\Finally, in Table~\ref{tab:orbitalExchangeJs} we compare the obtained theoretical values of the dominant orbiton hoppings with those extracted from fitting the orbital dispersion~\ref{orbitonDispersionNew}. We find relatively good agreement between theory and experiment, supporting the scenario discussed above. As a closing remark, we notice that in PNO also the third nearest-neighbor $t_3$ likely plays an important role in the orbiton propagation, as its extracted value is almost half of $t_1$. In the above discussion, we did not include $t_3$, as doing so would require going beyond the approximation used in the SM to solve the model. This constitutes a complex many-body problem that is beyond the scope of the present article. Nevertheless, the investigation of longer-range orbiton propagation dynamics is the next step in the refinement process of the model.
}
To summarize, our comparative study of the spin, charge and orbital excitations in IL nickelates and cuprates uncovers key similarities and distinctions that shed new light on their correlated electronic states. In IL nickelates, in-plane hopping integrals are smaller than in cuprates, leading to approximately half the nearest neighbor exchange interaction and to a substantially lower energy of the $d_{xy}$ excitation. However, in both families the absence of apical oxygens causes significant long-range hopping integrals, that shape the magnon in-plane dispersion and give rise to dispersing orbital excitations. The inter-plane direct exchange interaction seems little affected by the presence of a rare earth as in IL nickelates or of an alkali metal ion as in cuprates. In both cases, the out-of-plane coupling generates 3D \textcolor{black}{antiferromagnetic correlations} in the IL compounds. The rare-earth in the nickelates leads to a more tridimensional electronic structure and to self-doping effects resulting in a charge continuum that merges with $dd$ excitations in RIXS spectra. The orbital dispersion in the IL nickelate seems to primarily involve nearest neighbor orbital superexchange interaction, which in cuprates is strongly hampered by coupling to magnons. Combining these observations, we can speculate that IL nickelates are indeed mimicking the essential physics of cuprates, including the mechanisms for superconducting pairing. The lower $T_\text{c}$ in IL nickelates is mainly due to the overall smaller energy of spin fluctuations, which can be traced back to larger charge transfer $\Delta$ and/or smaller hopping integrals. In addition, the Mott-Hubbard nature of the correlation in IL nickelates, that cause a stronger localization of the doping charge on the metal site, might imply also a smaller electron-phonon interaction and the lack of charge density waves and fluctuations ubiquitous in hole doped cuprates. 
\\In conclusion, infinite-layer nickelates \textcolor{black}{appear to} share many key properties with cuprates, although with reduced interaction energies. \textcolor{black}{We can safely state that the exchange interaction between different planes clearly confers a certain degree of tridimensionality to the magnetism of PNO, notwithstanding the difficulty in exactly quantifying this effect through the size of the gap at $\Gamma$.} At the same time, the presence of the rare-earth ion enables a doping mechanism that is absent in infinite-layer cuprates because of the lack of apical oxygens. Together, these observations indicate that infinite-layer nickelates are unconventional superconductors closely related to cuprates, but lacking some of the ingredients that enhance $T_\text{c}$ in the latter.

\begin{table}
     \begin{tabular}{m{3.1cm} m{2.5cm} m{2.5cm}}
     \hline
     \hline

      Parameter (eV) &   PNO    & CCO \\
      \hline
       $t_{pd\sigma}$      &   1.04  & 1.30  \\
       $t_{pd\pi}$      &   0.56    &  0.70 \\
       $t_{pp}$      &   0.56    & 0.70  \\
       $\Delta$    &    \textcolor{black}{4.4}      &     1.80           \\
       $\epsilon_{\pi\sigma}$    &  -1.00     &    -1.60    \\
       $\epsilon_{xy}$    &   0.70    &      1.00  \\
       $U$    &   \textcolor{black}{6.5}    &     8.00   \\
       $J_H$    &  \textcolor{black}{1.2}     &   1.00     \\
       \hline
       \hline
    \end{tabular}
    \caption{\textbf{Microscopic parameters.} Charge transfer model parameters (in eV) employed for the calculation of $J_{1}^{orb}$ and $J_{2}^{orb}$ (see text for the model details).}
    \label{tab:modelParameters}
\end{table}

\begin{table}
     \begin{tabular}{m{3.2cm} m{2.8cm} m{2cm}}
        \hline
        \hline
        $J^{orb}$, $t^{orb}$ (meV) &   PNO    & CCO \\
       \hline
        Experiment $t_{1}^{orb}$   &  \textcolor{black}{$12.5\pm3.8$}  & -
        \\Experiment $t_{2}^{orb}$   &  \textcolor{black}{$0.2\pm4.8$}  & \textcolor{black}{$-13.6\pm3.5$}  \\
        Theory $J_{1}^{orb}$   &    \textcolor{black}{21.6}    &  -
        \\Theory $J_{2}^{orb}$   &   $-0.3$  &  $-15.0$   \\
        Theory $t_{1}^{orb}$   &   $13.7 $  &  -
        \\Theory $t_{2}^{orb}$   &   $\sim 0$  &  $-15$   \\
       \hline
       \hline
    \end{tabular}
    \caption{\textcolor{black}{\textbf{Orbital superexchange and orbiton hopping values.} Calculated values of the nearest- and next-nearest-neighbor orbital superexchange $J^{orb}$ (in meV) for PNO and CCO, together with the corresponding effective orbiton hopping amplitudes $t^{orb}$ (in meV), compared to those extracted from fits to the data in Fig.~\ref{orbitonsFig}(a)–(c); see text for further details.}
    }
    \label{tab:orbitalExchangeJs}
\end{table}

\section{Methods}
Pristine perovskite PrNiO$_3$ thin films were grown by RHEED-monitored Pulsed Laser Deposition on a SrTiO$_3$ (001) substrate (STO), up to a thickness of 15-20 unit cells ($\sim5$ nm), and capped with an epitaxial STO layer grown in-situ up to 12 unit cells ($\sim3$ nm). Superconductive IL phase PrNiO$_2$ was then obtained via a topotactic chemical reduction, consisting in the deintercalation of apical oxygen atoms by means of a high-$T$ annealing in hydrogen-rich atmosphere issued from a CaH$_2$ powder, as done in previous studies  \cite{lee2020aspects,Krieger2023growth,puphal2023synthesis,hayward1999sodium,hayward2003synthesis}. The resulting samples show a complete superconducting behavior below $T_\text{c} = 4$ K, while the SC onset (maximum curvature in the resistivity $\rho(T)$) is located slightly below $11$ K. The lattice parameters for the IL phase were found to be $a=3.94$ \AA\, and $c=3.28$ \AA. More details about sample growth and characterization are available in Ref. \onlinecite{sahib2025superconductivity}. 
\textcolor{black}{The CCO presented data are a re-analysis of the same raw dataset already partly published in Refs. \cite{martinelli2024collective,martinelli2022fractional}.} The CCO films were grown by pulsed-laser deposition (KrF excimer laser, $\lambda = 248$ nm) at a temperature around 600$^\circ$C and an oxygen pressure of $0.1$ mbar, on NdGaO$_3$ (NGO) (110) substrate. The substrate holder was at a distance of $2.5$ cm from the CCO target, which was prepared by a standard solid-state reaction  \cite{stellino2025role,di2015high}. The lattice constants for CCO, as determined from XRD, are $a=b=3.863$ \AA\, and $c=3.184$ \AA. \textcolor{black}{The measured resistivity of CCO at room temperature (300 K) was $\sim 10^4$ m$\Omega\cdot$cm.} Below 250 K, the film resistance exceeded the upper limit of our measurement setup \cite{martinelli2024collective}.
\\RIXS measurements were performed with the ERIXS spectrometer at the  ID32 beamline of the European Synchrotron (ESRF), in Grenoble  \cite{brookes2018beamline}. The beamline and spectrometer combined resolution was $40$ meV at the Ni L$_3$ edge ($852.4$ eV) of PNO and $42$ meV at the Cu L$_3$ edge ($930.6$ eV) of CCO. We measured momentum-resolved RIXS maps on both materials along the $M-\Gamma-X-M$ cuts of the BZ; for PNO, the Ni-$L_3$ resonance energy is actually not enough to reach the $X$ point, so we stopped at $\textbf{Q}=(0.475,0)$ (see the inset of Figure \ref{PNOCCOdata}(d)).
We used $\pi$ linear incident polarization and near grazing emission geometry (positive values of $H$), to enhance the spin non-conserving cross-section versus the spin conserving ones  \cite{braicovich2010momentum,ament2009theoretical,haverkort2010theory}. Selected spectra were taken with polarization resolution of the scattered photons  \cite{braicovich2014simultaneous,fumagalli2019polarization,brookes2018beamline}. In this case we employed both grazing incidence ($\theta=24^\circ$ as measured from the sample surface) and grazing emission ($\theta=125^\circ$) geometry, both with $\pi$ and $\sigma$ incident polarizations. All data were collected at $T = 20$ K with a fixed scattering angle $2\theta=149.5^\circ$; therefore the value of the out-of-plane momentum component $L$ was not fixed when scanning the in plane components $H$ and $K$. This is not an issue, since the in-plane exchange $J$s are still dominating the out-of-plane $J_{\perp}$.
\\All fittings were carried out with the software Matlab. The library SpinW \cite{spinWsite} was used for the extraction of the exchange couplings $J_x$ from the magnetic dispersion relations. For the orbital dispersion, calculation of theoretical $J^{orb}_{1}$ and $J^{orb}_{2}$ was based on the same charge transfer model reported in the Supplementary Information of Reference \onlinecite{martinelli2024collective}.
\\\textcolor{black}{Concerning the parameter choice for the orbiton propagation model, for PNO the hopping integrals were reduced to 80\% of their corresponding CCO values, to account for the larger $d$–$p$ energy splitting \cite{botana2020similarities}.
The splittings $\epsilon_{\pi\sigma}$ and $\epsilon_{xy}$ values \textcolor{black} were taken from electronic structure calculations \cite{botana2020similarities}. 
The value of Hund's exchange was estimated based on the atomic value for the nickel ion~\cite{Haverkort2005}, assuming a significantly smaller reduction in the solid state than that typically observed in cuprates. This is attributed to the reduced nephelauxetic effect (orbital cloud expansion)~\cite{schaffer1958nephelauxetic} in nickelates, where lower covalence preserves the intra-atomic nature of the exchange interaction more effectively than in the highly covalent cuprates.
\\\textcolor{black}{Since there is no clear consensus in the literature on the values of the charge-transfer energy $\Delta$ and the Mott interaction $U$, we adopted $\Delta$ from DFT calculations \cite{botana2020similarities}, while $U$ was taken from values proposed for the trilayer nickelate La$_4$Ni$_3$O$_8$ \cite{shen2022role}. Although the Ni valence in that compound is not identical to that of the infinite-layer nickelates, this choice is consistent with the mixed Mott–Hubbard/charge-transfer regime proposed as a common framework for the nickelate family \cite{shen2023electronic}. The $\Delta$ value has also been very recently corroborated by photoemission measurements on the $n=8$ nickelate Nd$_9$Ni$_8$O$_{18}$ \cite{scott2026contrasting}.}}

\section{Data availability}
All data shown in the main text and in the Supplementary Information are available at the Zenodo repository at \textcolor{blue}{https://zenodo.org/records/21440348}, DOI: 10.15151/ESRF-ES-1430231833.

\section{Supplementary Information}
\label{SupplementMater}

\subsection{Absence of charge order}
The quasi elastic integrals of the $[H;0]$ spectra of PNO are reported in figure \ref{noChargeOrder}. No detectable intensity increase is present around the wavevector $\textbf{Q}=[1/3;0]$, where charge order in IL nickelates is supposed to appear\cite{rossi2022broken,krieger2022charge,parzyck2024absence}.

\subsection{RIXS spectra fitting}
Low-energy RIXS fittings reported in Figure 2(a-b) in the main text were carried out by employing two Gaussian functions for the elastic and phonon peak. \textcolor{black}{Phonon and magnon overtones  were considered only in the cuprate fit, since in the nickelate these features are too broad and too close in energy to be reliably disentangled. The phonon was removed from the points closest to $\Gamma$, both along $(H,0)$ and $(H,H)$. In this region, the elastic intensity increases near the specular condition, while the magnon softens to very low energy, making a reliable disentanglement of magnon and phonon contributions difficult. Since the measurements were performed in $\pi$ geometry, which emphasizes magnetic excitations, including the phonon in the fit could lead to an overestimation of its weight and hence compromise the extraction of the magnon energy. The quality of the fits near $\Gamma$ remains unaffected, as confirmed by the fitted spectra.}
\\As explained \textcolor{black}{in the main text}, the magnon was fitted with a damped harmonic oscillator function \cite{peng2018dispersion}: we consider that here a single-peak fit is accurate enough for our scopes, although for CCO it results in an overestimation of the single-magnon damping energy close to (0.5,0). In PNO, another broad Gaussian on the right accounts for the tail of higher-energy features, mainly the lowest-energy $dd$ and the RE hybridization peak at 0.6 eV \cite{hepting2020electronic}. For CCO we also employed an antisymmetrized exponential starting at $0.3$ eV to take into account multiple magnetic excitations (bimagnons), and a Gaussian for the phonon overtone (biphonon). Such higher order peaks are not independently included in the PNO fit, because these features are too close in energy and too broad. Our anti-symmetrization process started from a 
simple Gaussian:

\begin{figure*}[]
\centering
\includegraphics[width=0.8\textwidth]{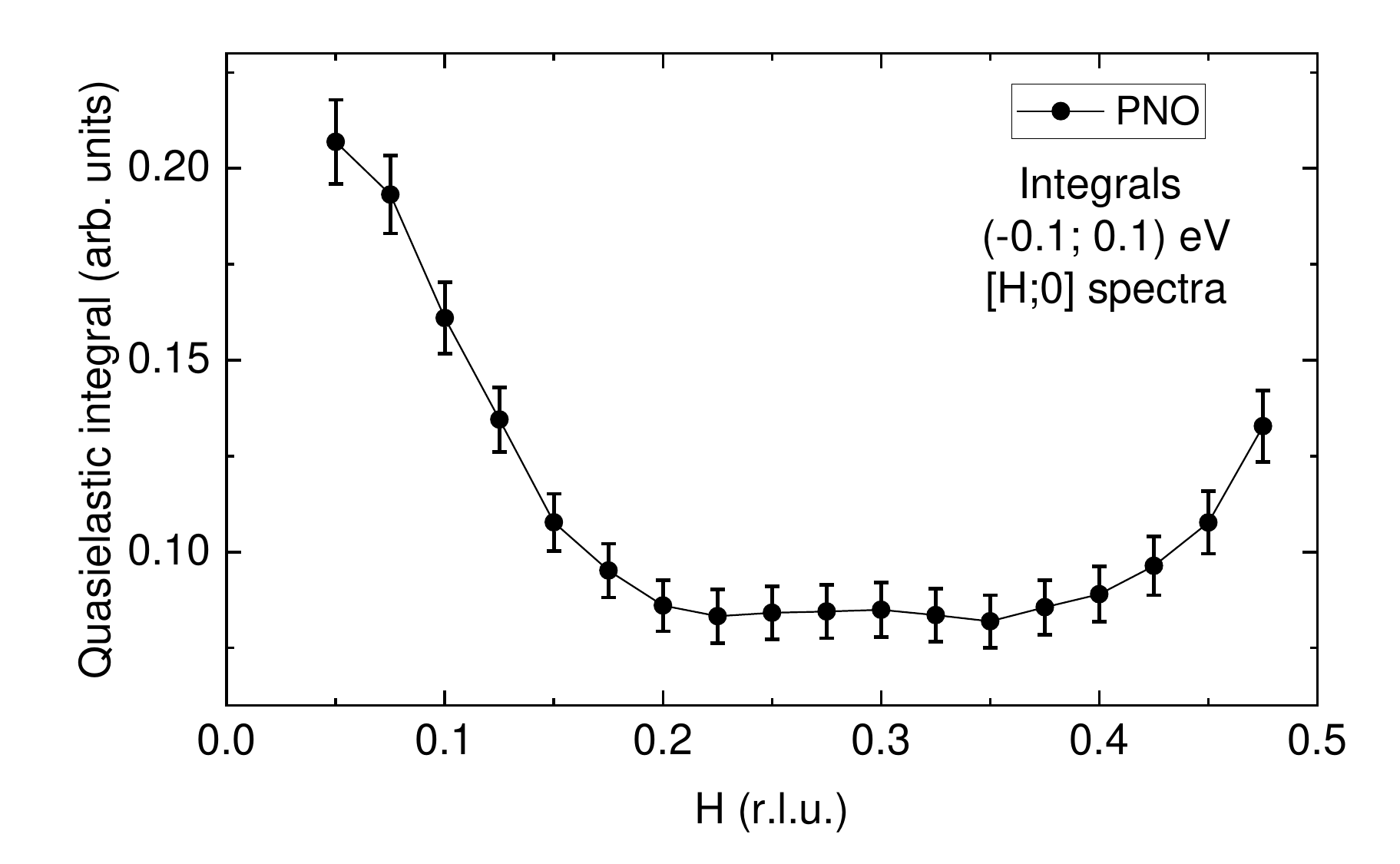}
\caption{\label{noChargeOrder} \textbf{Absence of charge order.} Integrals of $[H;0]$ spectra of PNO (figure 1(b) in the main text) on the quasielastic energy range $[-0.1;0-1]$ eV. Error bars are given from shot noise estimation in the experimental spectra, evaluated as $\sigma\,\propto \sqrt{\text{Single Photon Counting}}$.
} 
\end{figure*}

\begin{equation}
    G(I,E,\sigma,x)=Ie^{-\frac{x-E}{2\sigma^2}};
\end{equation}

where $E,I,\sigma$ are the fitting parameters and $x$ is the energy loss variable. A symmetric copy with respect to $x=0$ was then subtracted, and the resulting odd-symmetric function was then summed to its absolute value and divided by 2. As a result, we obtain a continuous curve , that is identically null for $x\leq0$:

\begin{equation}
    Y(I,E,\sigma,x)=\frac{G(I,E,\sigma,x)-G(I,-E,\sigma,x) + \bigg|G(I,E,\sigma,x)-G(I,-E,\sigma,x)\bigg|}{2}
\end{equation}

Finally, the curve was translated rightwards to make it start from $0.2-0.3$ eV, which is the typical energy for multiple magnetic excitations in cuprates. The precise starting point was fixed slightly differently for different spectra, according to the quality of the resulting fit.

\textcolor{black}{
\subsection{Orbital transitions fitting}
}

\begin{figure}[]
\centering
\includegraphics[width=\textwidth]{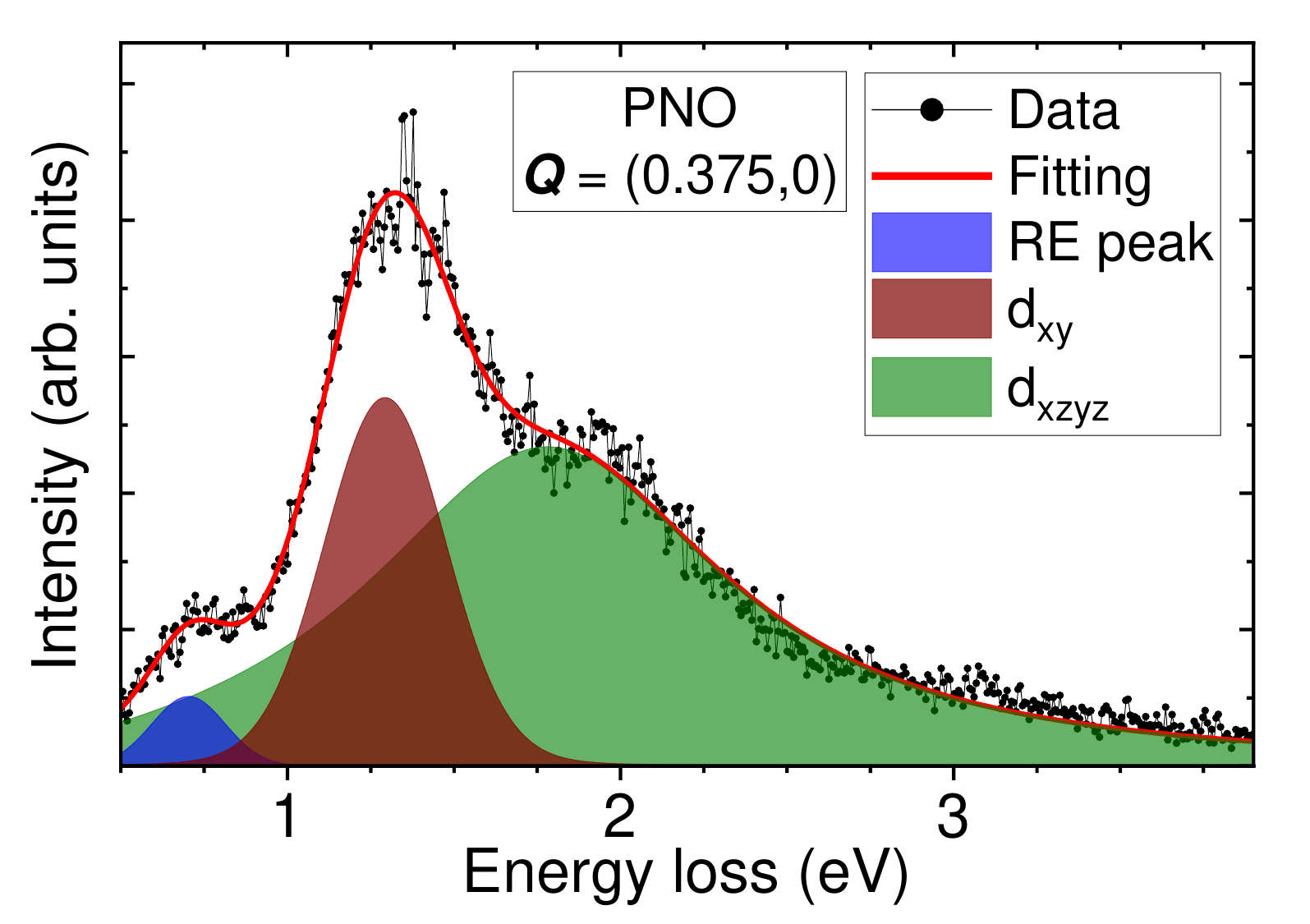}
\caption{\label{fitPNO_only2dds} \textcolor{black}{\textbf{PNO fitting with only two $dd$ peaks.} Fitting of RIXS spectrum at \textbf{Q}=(0.375,0) without the $3d_{z^2}$ peak.}} 
\end{figure}

For fittings of the $dd$ orbital peaks, Voigt profiles were employed (Figure 2(c-d) in the main text). \textcolor{black}{These were antisymmetrized according to the same procedure followed for the bimagnon continuum, and convolved with a Gaussian accounting for instrumental resolution. We employed three orbital peaks, consistently with expectations for a planar square crystal field environment \cite{sala2011energy,rossi2021orbital}. While the first two peaks are already quite evident from raw data, the fitting of the highest energy one, i.e., $3d_{z^2}$, is more delicate, as its large broadness makes it appear only as a mild shoulder at around 3 eV. As a different attempt, we also tried to fit the spectra by ignoring this contribution and using only two $dd$ peaks, as shown in Figure \ref{fitPNO_only2dds}. If compared to Figure 2(c) in the main text, the quality of the fitting is sensibly worsened in all the interval $\geq2$ eV, with a shoulder at 3 eV which is not captured by the $3d_{xz/yz}$ tail, thus justifying the employment of an extra peak.}

\begin{figure}[]
\centering
\includegraphics[width=0.8\textwidth]{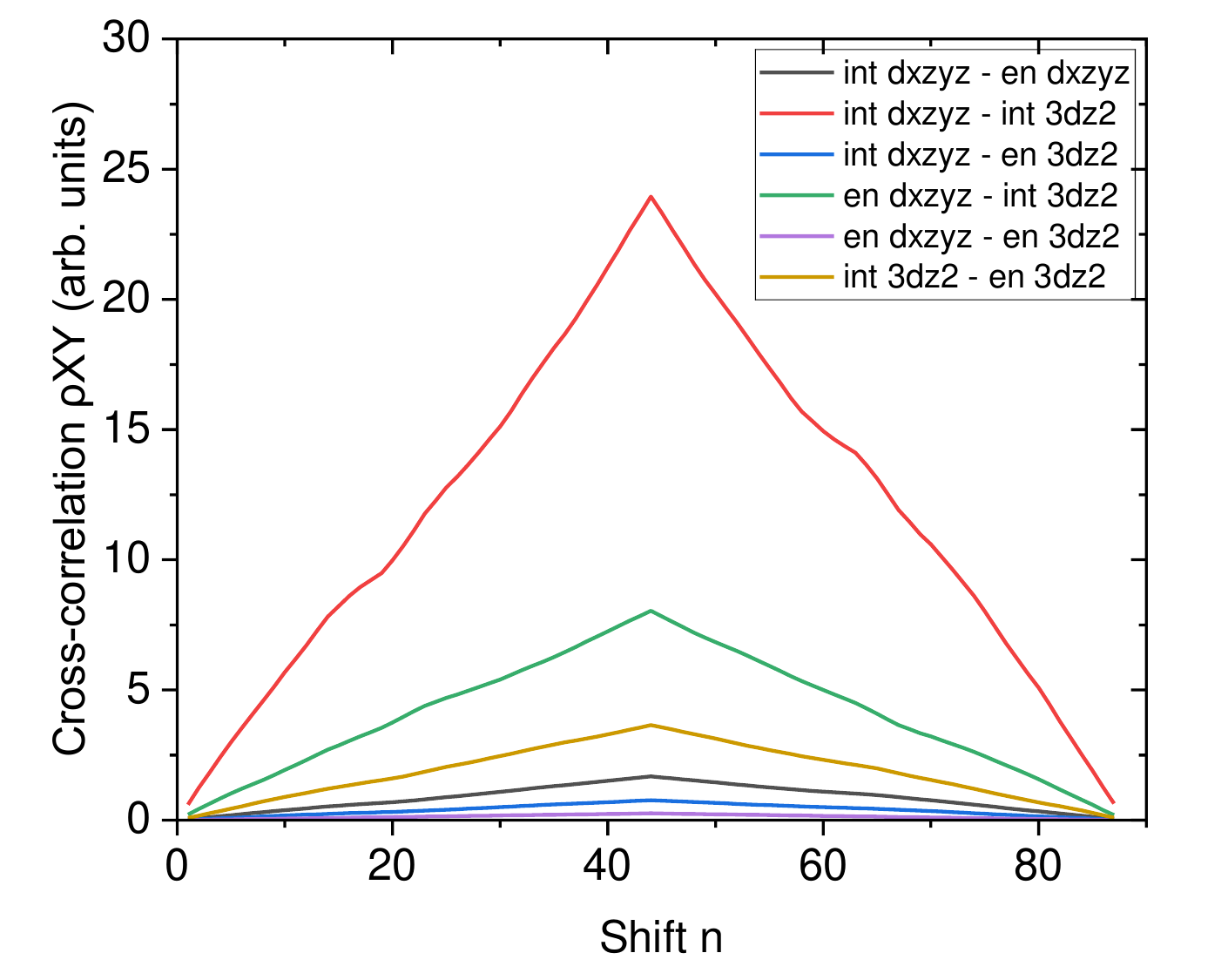}
\caption{\label{CrossCorr_dds_fig} \textcolor{black}{\textbf{Cross-correlation parameters for $dd$ fittings.} Cross-correlation between intensities and energies of the fitting Voigt functions for the two highest-energy $dd$ peaks.}} 
\end{figure}

\textcolor{black}{Cross-correlation parameters between each couple of energies and intensities of the $3d_{xz/yz}$ and $3d_{z^2}$ are displayed in Figure \ref{CrossCorr_dds_fig}. These were normalized to the product of the variances of the two parameters $X$ and $Y$ considered in each case, according to the formula \ref{CrossCorrelEquation}:}

\begin{equation}
\label{CrossCorrelEquation}
    \textcolor{black}{\rho_{XY}(n)=\frac{\sum_{i}{X_iY_{i+n}}}{\sigma_X\sigma_Y}}
\end{equation}

\textcolor{black}{with the index $i$ running over all the momenta of the measured Brillouin zone cuts.}

\begin{figure}[]
\centering
\includegraphics[width=\textwidth]{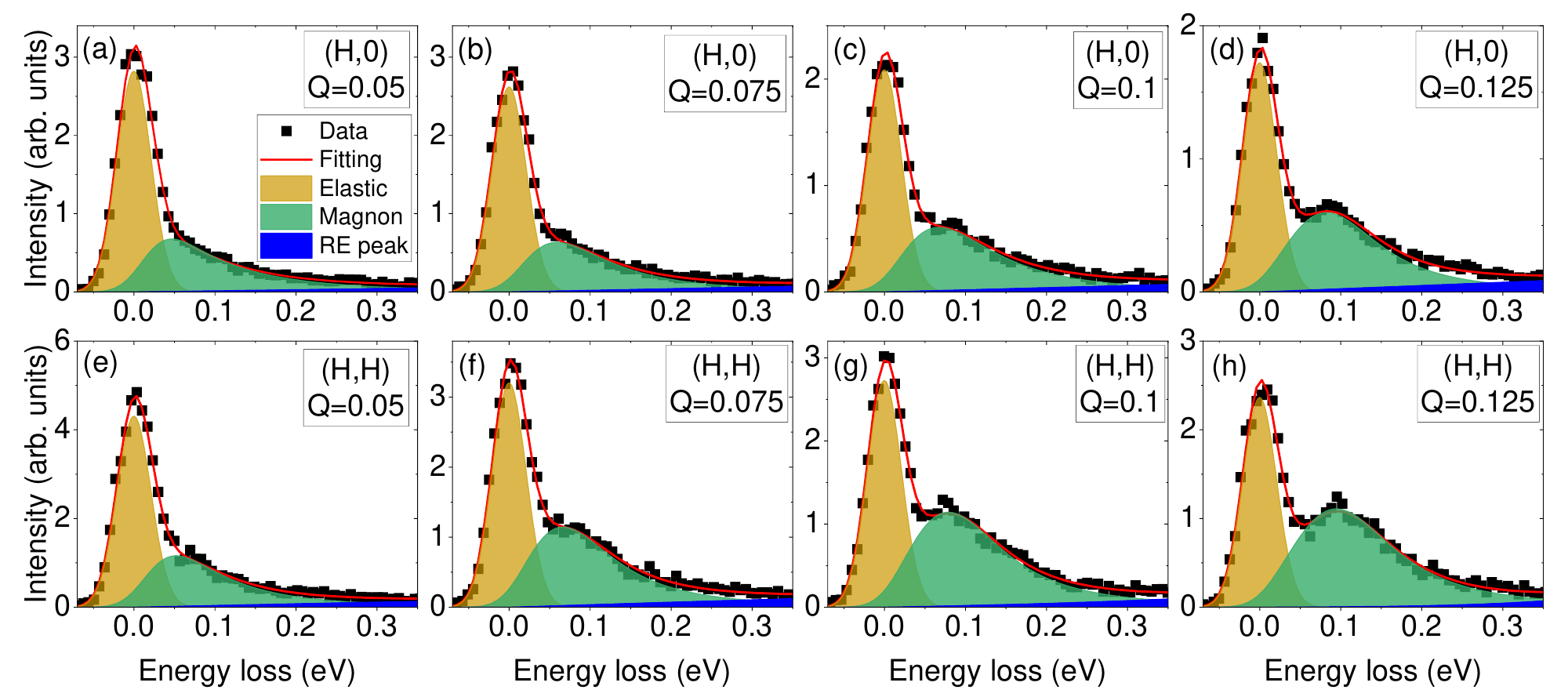}
\caption{\label{lowHfittings} \textcolor{black}{\textbf{Low-$H$ RIXS fittings.} Fittings of low-momentum RIXS spectra, along $(H,0)$ (a-d) and $(H,H)$ (e-h), at the in-plane momentum $Q$ reported inside each panel.}} 
\end{figure}


\begin{figure}[]
\centering
\includegraphics[width=\textwidth]{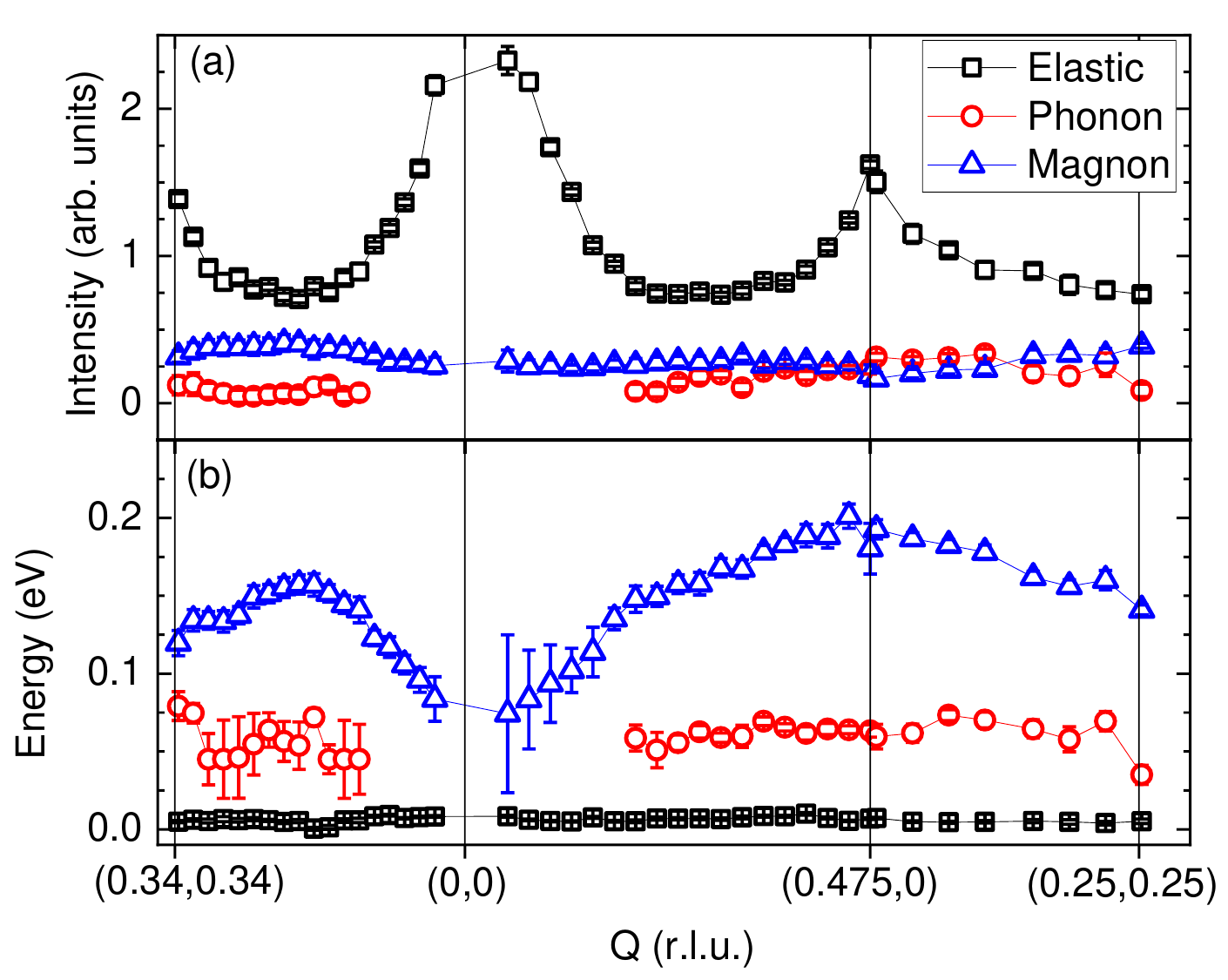}
\caption{\label{lowEnergy_trends_fig} \textcolor{black}{\textbf{Elastic, phonon and magnon fitting parameters.} Trends of the elastic (black symbols), phonon (red) and magnon (blue) intensities (a) and energies (b) in PNO as extracted from fittings.}} 
\end{figure}

\begin{figure}[]
\centering
\includegraphics[width=0.8\textwidth]{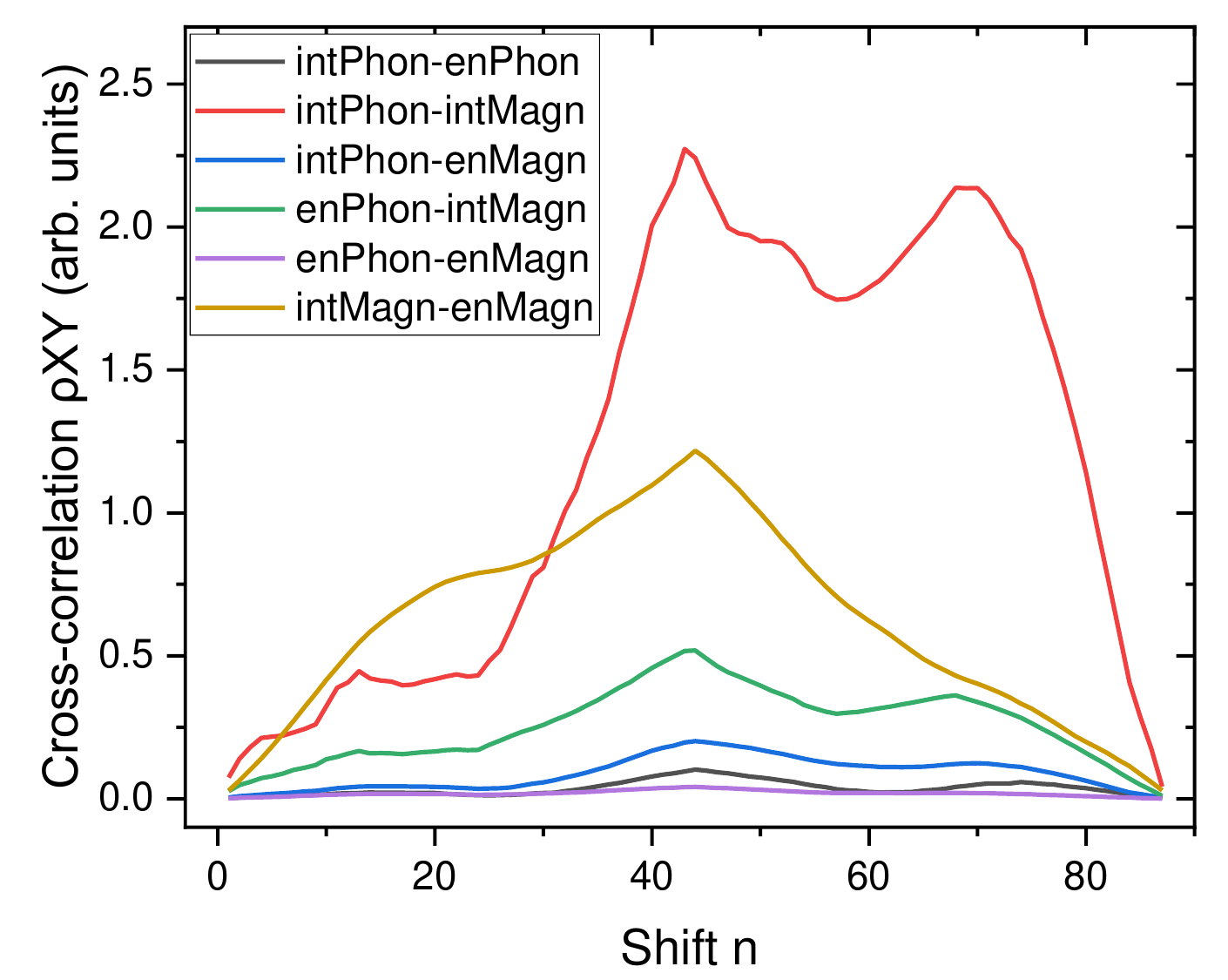}
\caption{\label{CrossCorr_lowEn_fig} \textcolor{black}{\textbf{Cross-correlation parameters for low-energy fittings.} Cross-correlation between intensities and energies of the elastic, phonon and magnon peak.}} 
\end{figure}

\textcolor{black}{
\subsection{Low-H fittings}
Fittings of RIXS spectra at low in-plane momentum are quite delicate, both because of the elastic increase close to the specular condition and of the low energy of the magnon, causing it to merge with the elastic. This is particularly relevant since it makes difficult to estimate the gap size at the $\Gamma$ point. At the same time, it makes difficult to safely fit low-energy excitations such as phonons, due to insufficient energy resolution. For this reason, we removed the phonon peak from the spectra closest to the $\Gamma$ point. More precisely, we discarded the phonon contribution for an in-plane momentum $Q$ up to 0.175 r.l.u. and 0.125 r.l.u. along $(H,0)$ and $(H,H)$ respectively. In Figure \ref{lowHfittings} we report the results of the corresponding fittings for the first four points closest to $\Gamma$ in each direction. It is seen that our choice does not compromise the quality of the fitting at low $Q$, as all the spectral weight can be properly fitted by the elastic and magnetic contributions alone.
 \\Figure \ref{lowEnergy_trends_fig} displays the intensity and energy trends of the elastic, phonon and magnetic peak (black, red and blue symbols respectively). The elastic increase close to $\Gamma$ is caused by the specular condition, while another elastic increase is seen for large in-plane momenta, which has already been observed, but not discussed, in several other samples (see main text). Cross-correlations between phonon and magnon intensities and energies, calculated according to Equation \ref{CrossCorrelEquation}, are displayed in Figure \ref{CrossCorr_lowEn_fig}.}

\begin{figure}[]
\centering
\includegraphics[width=0.8\textwidth]{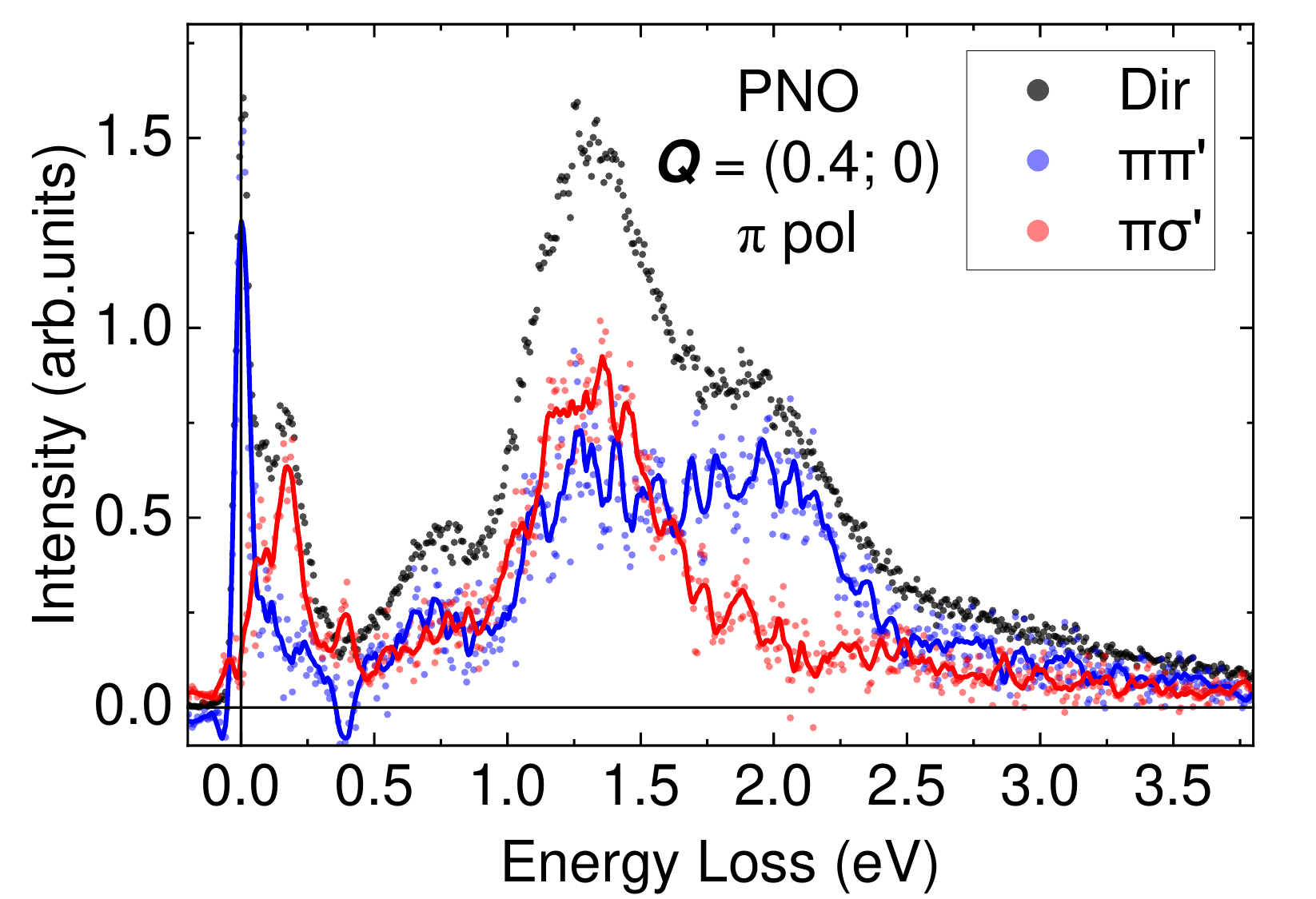}
\caption{\label{smoothPolarim_fig} \textcolor{black}{\textbf{PNO smoothed polarimetric RIXS spectrum.} Polarimetric spectrum of PNO with grazing emission geometry and $\pi$ incident polarization. Red and blue dots represent extracted crossed and non-crossed polarimetric components, while continuous lines display 15-points smoothings of the respective components.}} 
\end{figure}

\textcolor{black}{
\subsection{Polarimeter data smoothing}
Figure \ref{smoothPolarim_fig} reports PNO polarimetric data for grazing emission geometry and $\pi$ incident polarization (same as Figure 4(c) in the main text). To better visualize the polarized or non-polarized nature of the different peaks, we performed a 15-points smoothing of the two components, represented as continuous lines. It can be seen how the $\approx0.7$ eV peak indeed shows a mixed-nature, in contrast, for instance, with the magnon peak at $\approx0.2$ eV, which is almost completely polarized, except for the small non-crossed underlying background.
}

\begin{figure}[]
\centering
\includegraphics[width=0.8\textwidth]{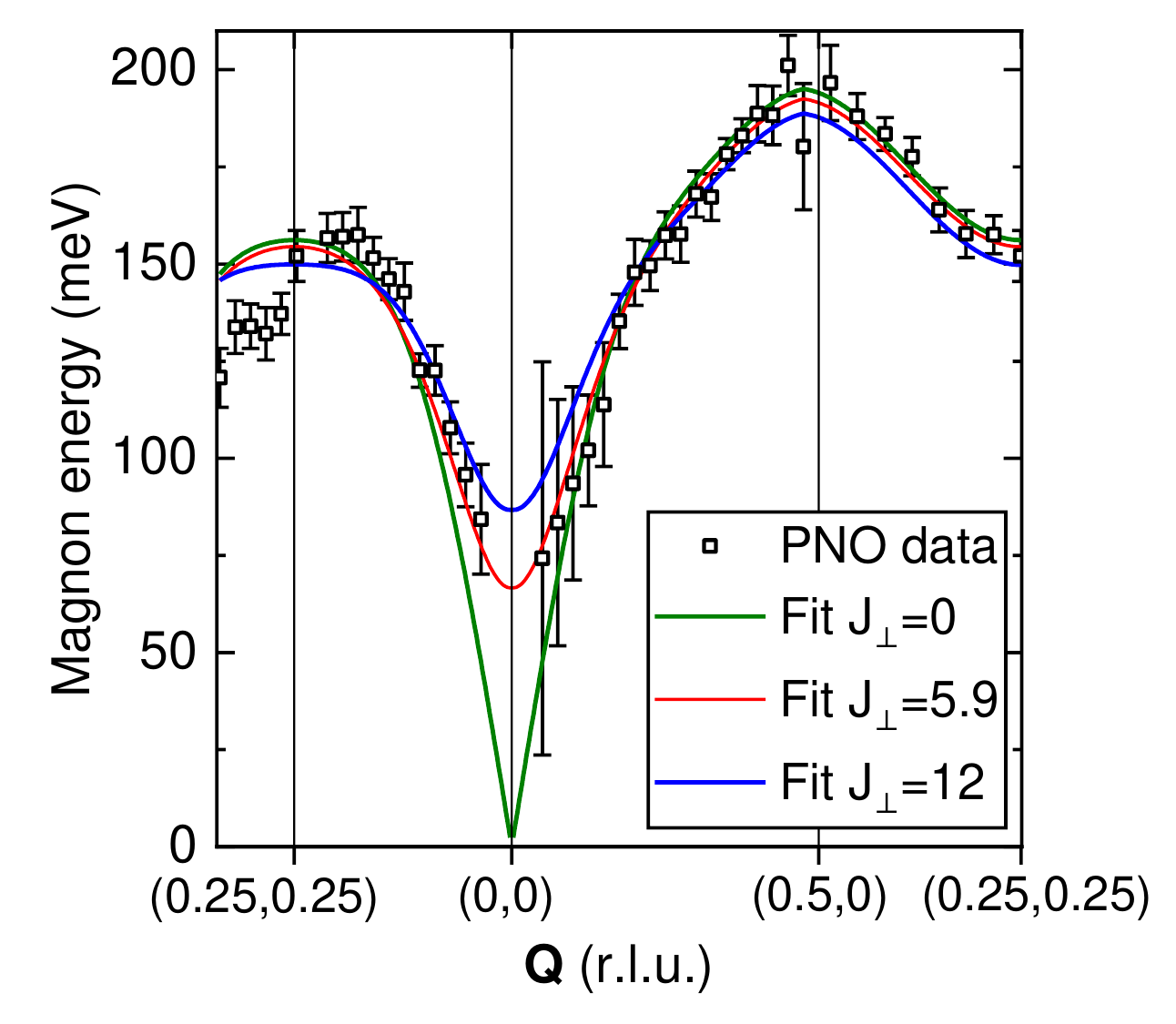}
\caption{\label{JperpPNO_fig} \textcolor{black}{\textbf{PNO $J_\perp$ estimation.} Comparison between LSW fittings of the PNO magnon dispersion for different $J_\perp$ values.}} 
\end{figure}

\textcolor{black}{
\subsection{Estimation of $J_\perp$ for PNO magnon dispersion}
As pointed out in the main text, the extraction of the $J_\perp$ value is problematic for PNO, due to the difficulty in safely estimating the value of the gap at the $\Gamma$ point. Figure \ref{JperpPNO_fig} reports the comparison between fitting attempts for different fixed $J_\perp$ values. Red curve represents the fitting with the $J_\perp$=5.9 meV value used in the main text. It is clear that ignoring the presence of the gap ($J_\perp$=0, green curve) sensibly worsens the fitting. On the other hand, a value of 12 meV (blue) results already too large. We can therefore reasonably estimate an interlayer exchange coupling $J_\perp$ between 0 and 10 meV, not too different from that of CCO.
}

\textcolor{black}{
\subsection{Single-ion RIXS cross-section calculations}
Single-ion calculations were performed following the approach described in Ref. \cite{sala2011energy}. The spin–orbit coupling was set to zero. The spin direction was assumed to lie within the $ab$ plane and oriented at 45$^\circ$ with respect to the $a$ axis, for both CCO and PNO systems. Each excitation peak was modeled as the sum of two contributions corresponding to non-spin-flip (nsf) and spin-flip (sf) channels. For example, the total intensity of the $z^2$ excitation is given by: $A_{z^2}=A_{z^2}(nsf)+A_{z^2}(sf)$. Here, “nsf” refers to processes in which the $z^2$ electron filling the core hole has the same spin as the electron added to the $x^2-y^2$ state, while “sf” denotes the opposite spin configuration. 
All $dd$ excitations were modeled using Lorentzian lineshapes. For PNO, the full width at half maximum (FWHM) values were set to: 1.2 eV for the $z^2$ excitation, 0.4 eV for the $xy$ excitation, and 0.8 eV for the $xz/yz$ excitations. In this case, the nsf and sf components were assigned the same broadening.
For CCO, the FWHM values were: 0.38 eV for $z^2$ and 0.16 eV for $xy$, again with identical broadenings for the nsf and sf components. The $xz$ and $yz$ features were treated separately to reproduce the remaining peaks.  A better agreement with the experimental spectra was obtained by modeling these excitations as the sum of two Lorentzian components (nsf and sf) with different broadenings. In particular, the corresponding FWHM values were: $xz_{nsf}$= 0.2 eV, $xz_{sf}$ = 0.1 eV, $yz_{nsf}$ = 0.54 eV and $yz_{sf}$ = 0.2 eV.
}

\textcolor{black}{
\subsection{Mott energy $U$ extraction}
It must be mentioned that our estimate of the Mott energy $U$ value is exclusively based on the orbital dispersion, thus it can be significantly different from other magnon-derived estimates \cite{rosa2024spin}.
}

\textcolor{black}{
\subsection{Spin-orbital model}
\subsubsection{The Hamiltonian}
As the starting point, we follow results from the main text which show that in PNO solely
the nearest-neighbor superexchanges matter -- both for the spin as well as for the spin-orbital case. This leads us to formulate the following (Kugel-Khomskii) spin-orbital Hamiltonian: 
\begin{align}\label{eq:H}
H=&H_{\rm spin} + H_{\rm xy}, \\
H_{\rm spin} =& {J_1} \sum_{\langle i, j \rangle || \hat{x}, \hat{y}} \mathcal{P}_{i,j} \left( {\bf S}_i {\bf S}_j - \frac14 \right), \\ 
H_{\rm xy}=& J_{\rm HS} \sum_{\langle i, j \rangle || \hat{x}, \hat{y}}   \left( \tau^+_i \tau^-_j  + \tau^-_i \tau^+_j\right) \left( {\bf S}_i {\bf S}_j +\frac34 \right) 
- J_{\rm LS} \sum_{\langle i, j \rangle || \hat{x}, \hat{y}}   \left( \tau^+_i \tau^-_j  + \tau^-_i \tau^+_j\right) \left( - {\bf S}_i {\bf S}_j +\frac14 \right) \nonumber \\
&+ E_{\rm xy} \sum_i \tau^z_i,
\end{align}
where $H_{\rm spin}$ is the well-known Heisenberg spin exchange between the $S=1/2$ spins present on all bonds without orbital excitations
and $H_{\rm xy}$ denotes the spin-orbital superexchange which is responsible for the (correlated) hopping 
of an orbital excitation with an on-site energy cost equal to $E_{xy}$. Whereas the spin operators ${\bf S}_i$ are defined in a usual way,
the orbital operators ${\bf \tau}_i$ need to be explicitly defined to avoid ambiguity: $\tau^z_i = (\tilde{n}_{i, xy} - \tilde{n}_{i, x^2-y^2})/2$,
$\tau^+_i = \tilde{c}^\dag_{i, xy} \tilde{c}_{i, x^2-y^2}$, $\tau^-_i = \tilde{c}^\dag_{i, x^2-y^2} \tilde{c}_{i, xy}$ with the constrained fermion
operators $\tilde{c}^\dag_{i, \alpha}$ ($\tilde{c}_{i, \alpha}$) creating (annihilating) fermions on site $i$ and in orbital $\alpha$ and acting
in the restricted Hilbert space without doubly occupied sites. Finally, the operator $\mathcal{P}_{i,j} $ ensures that there are no orbital
excitations along the spin-only exchange bonds.
Note that this form of the spin-orbital Hamiltonian is a generic one for the case of an orbital excitation moving in the antiferromagnetic and ferroorbital background and follows from the detailed considerations that can be found e.g. in~\cite{Wohlfeld2013}.
\\There are four parameters of the model Hamiltonian: 
\begin{enumerate}
\item Spin exchange $J_1$ which is of the order of $46$ meV in PNO (see main text). 
\item Two spin-orbital superexchange parameters $J_{\rm HS} \equiv j_{\rm HS} \bar{J}^{orb}_{1} $ and $J_{\rm LS} \equiv j_{\rm LS} \bar{J}^{orb}_{1}$ 
with $j_{\rm HS} = 1/(1-3J_H/U) = 2.24$ and $j_{\rm LS} = 1/(1-J_H/U)  = 1.23$ by assuming that the Hund's exchange $J_H = 1.2$ eV and $U=6.5$ (see Table II of the main text and the accompanying discussion), and $ \bar{J}^{orb}_{1} = U J^{orb}_{1} /(U - 2J_H)\approx 27 $ meV (see main text for the estimation of $J^{orb}_{1}$, i.e. the average value of the orbital exchange for two antiparallel spins; note that $\bar{J}^{orb}_{1}$ is defined as an `auxiliary' orbital exchange, with the intermediate state costing $U$). 
\item $E_{\rm xy}$ is the on-site energy cost of the orbital excitation -- its value is determined from the PNO
RIXS spectrum and is equal to 1.29 eV (see main text).
\end{enumerate}
Finally, we also define  the orbital dynamical structure factor that decribes the dynamics of a single orbital excitation added to the antiferromagnetic and ferroorobital ground state of model $H$ and whose motion is governed by the spin-orbital Hamiltonian $H$:
\begin{align} \label{eq:Okw}
O_{\rm xy} ({\bf k}, \omega)= - \frac{1}{\pi} {\rm Im}   \langle GS | \tau^-_{\bf k} \frac{1}{\omega -H +E_{GS} + i 0} \tau^+_{\bf k} |GS \rangle.
\end{align}
Here $|GS \rangle$ is the (ferroorbital and antiferromagnetic) ground state of $H$ with energy $E_{GS}$ ($N$ is the number of the lattice sites in the system).
}
\textcolor{black}{
\subsubsection{Mapping onto a polaronic model}
In the next step we map the orbital problem defined by Eqs.~(\ref{eq:H},\ref{eq:Okw}) onto a polaronic problem we perform the following transformations:
\begin{enumerate}
\item We rotate all spins on one sublattice (in the usual way, see e.g. Ref.~\cite{Martinez1991} -- so that the antiferromagnetic ground state becomes ferromagnetic
with all spins pointing up).
\item We introduce the Holstein-Primakoff transformation for spins and pseudospins (anticipating the linear spin/orbital wave approximation below,
we skip the `square roots')  :
\begin{align}
S^z_i = \frac12 - n_{i \alpha}, \quad S^+_i = \alpha_i, \quad S^-_i = \alpha_i^\dag, \\
\tau^z_i =  n_{i \beta} - \frac12, \quad \tau^+_i = \beta_i^\dag, \quad \tau^-_i = \beta_i,
\end{align}
with $\alpha$ ($\beta$) being the boson operators that are called magnons (orbitons), respectively.
\item We introduce the hard-core boson operators, denoting two types of orbitons ($h_{i \downarrow}$ and $h_{i \uparrow}$), in the following manner
\begin{align}
h^\dag_{i \uparrow} = \alpha^\dag_i \beta^\dag_i, \quad h^\dag_{i \downarrow} = \beta^\dag_i (1-n_{i \alpha}).
\end{align}
\item We keep only: (i) up to quadratic terms in magnons $\alpha_i$, (ii) up to  quadratic terms in orbitons  $h_{i \uparrow}$ and $h_{i \downarrow}$, (iii) up to quadratic and linear terms in the mixed orbiton-magnon
terms (up to quadratic in orbitons and linear in magnons).
\item We perform successive Fourier and Bogoliubov transformation, see e.g. Ref.~\cite{Martinez1991} (the latter is performed solely for the magnons and introduces the `Bogoliubov magnons' $\bar{\alpha}_{\bf k}$). 
\end{enumerate}
Altogether, we obtain the following polaronic Hamiltonian
\begin{align}\label{eq:Hpol}
\bar{H}&=\bar{H}_{\rm spin} + \bar{H}_{\rm xy}, \\
\bar{H}_{\rm spin}& = \frac{zJ_1}{2} \sum_{ {\bf k }}  \sqrt{1-\gamma_{\bf k}^2}  \left( \bar{\alpha}^\dag_{\bf k}  \bar{\alpha}_{\bf k} + \frac12 \right), \\ 
\bar{H}_{\rm xy}&= E_{\rm xy} \sum_{\bf k} \left( h^\dag_{{\bf k}, \downarrow} h_{{\bf k}, \downarrow} +  h^\dag_{{\bf k}, \uparrow} h_{{\bf k}, \uparrow} \right)
+ \frac{j_{\rm HS} - j_{\rm LS}}{2} z \bar{J}^{ orb}_{1} \sum_{\bf k} 
\label{eq:Hpolmain} \gamma_{\bf k} h^\dag_{{\bf k}, \downarrow} h_{{\bf k}, \downarrow} \\
& + \frac{j_{\rm HS} + j_{\rm LS}}{2} \frac{z\bar{J}^{ orb}_{1}}{\sqrt{N}}  \sum_{{\bf k}, {\bf p}}
\Big[ h^\dag_{{\bf k}, \downarrow} h_{{{\bf k}-{\bf p}}, \uparrow} \bar{\alpha}_{\bf p} \left(u_{\bf p} \gamma_{{\bf k} - {\bf p}} + v_{\bf p} \gamma_{\bf k} \right) + h.c. \Big] \\
& + j_{\rm HS} \frac{z\bar{J}^{ orb}_{1}}{\sqrt{N}}  \sum_{{\bf k}, {\bf p}}
\Big[ h^\dag_{{\bf k}, \uparrow} h_{{{\bf k}-{\bf p}}, \downarrow} \bar{\alpha}_{\bf p} \left(u_{\bf p} \gamma_{{\bf k} - {\bf p}} + v_{\bf p} \gamma_{\bf k} \right) + h.c. \Big],
\end{align}
\\where $z=4$, $\gamma_{\bf k} = (\cos k_x + \cos k_y)/2$, and $u_{\bf k}, v_{\bf k}$ are the so-called Bogoliubov factors defined in the usual way.   
\\The orbital dynamical structure factors can now be approximated by
\begin{align} \label{eq:Okwpol}
\bar{O}_{\rm xy} ({\bf k}, \omega)= - \frac{1}{\pi} {\rm Im}   \langle \bar{GS} | h_{{\bf k}, \downarrow} \frac{1}{\omega -\bar{H} +E_{\bar{GS}} + i 0} h^\dag_{{\bf k}, \downarrow}  |\bar{GS} \rangle.
\end{align}
}

\textcolor{black}{
\subsubsection{Approximate orbital dynamical structure factor}
Finally, our main goal is to simplify the above Hamiltonian and this way be able to efficiently calculate the orbital dynamical structure.
To this end, we note that the coupling orbitons and magnons in Eq.~\eqref{eq:Hpol}
may only lead to a very weak orbiton bandwidth, due to the very strong renormalisation of the orbiton bandwidth after coupling to magnons~\cite{Martinez1991}, {\it cf}. discussion and results in \cite{martinelli2024collective} and~\footnote{We note that, on top of a large incoherent spectrum due to the orbiton scattering on magnons, such a coupling may lead to a small but finite effective
orbiton hopping not solely between NN but also between NNN or even third neighbors~\cite{Martinez1991}.}
Hence, we skip the latter terms and obtain a completely `free' theory -- i.e. with solely the free orbiton hopping term 
(\ref{eq:Hpolmain}). This way we obtain that the orbital dynamical structure factor is proportional to Dirac delta function 
\begin{align}
    \bar{O}_{\rm xy} ({\bf k}, \omega) \approx \delta (\omega - \varepsilon^{orb}_{\bf k} )
\end{align}
that is peaked at the orbiton dispersion relation given by the nearest neighbor dispersion relation
\begin{align} \label{eq:varepsilon^orb}  \varepsilon^{orb}_{\bf k} =   E_{\rm xy} +  z t^{orb}_{1} \gamma_{\bf k}. \quad {\rm where} \quad t^{orb}_{1} = \frac{j_{\rm HS} - j_{\rm LS}}{2} \bar{J}^{ orb}_{1}. \end{align}
Crucially, plugging in the parameters mentioned above, we obtain 
$t^{{orb}}_{1} = 13.7$ meV for PNO---i.e. ar rather large value, that is in good agreement with the one observed experimentally (see main text for more detals). }

\textcolor{black}{Interestingly, one may also ask what happens if a similar analysis is performed for CCO. Here, the situation is somewhat subtle. As discussed in the Supplementary Information of Ref.~\cite{martinelli2024collective}, the nearest-neighbor orbital exchange is actually finite, with $J^{\mathrm{orb}}_1 = 22$ meV. However, this contribution was entirely neglected in the analysis of Ref.~\cite{martinelli2024collective}, where it was argued---just as in the main text of the present paper---that this exchange primarily gives rise to orbiton scattering on magnons. Moreover, the very large NNN orbital exchange in CCO allows for a free-orbiton propagation channel, which constitutes a second, equally important, reason why the NN orbital exchange was neglected in Ref.~\cite{martinelli2024collective}.}

\textcolor{black}{Nevertheless, the situation encountered above for PNO suggests that a finite NN orbiton hopping amplitude, $t^{\mathrm{orb}}_1$, could also arise in CCO due to the above-mentioned free orbiton hopping mechanism enabled by a finite Hund's exchange. To investigate this possibility, we insert the CCO parameters used in Ref.~\cite{martinelli2024collective} into Eq.~\eqref{eq:varepsilon^orb} and evaluate $t^{\mathrm{orb}}_1$ for CCO. We find that this value is substantially smaller in CCO than in PNO, namely $t^{\mathrm{orb}}_1 = 6.7$ meV -- which, together with a very large NNN orbiton hopping in CCO, largely justifies why one could at first order neglect this hopping in the analysis of~\cite{martinelli2024collective}. The above reduction in the NN orbiton hopping in CCO originates from the smaller Hund's exchange, $J_H = 1$ eV, and the larger Hubbard interaction, $U = 8$ eV, adopted for CCO in Ref.~\cite{martinelli2024collective}, compared with the values used here for PNO ($J_H = 1.2$ eV and $U = 6.5$ eV; see the main text and below).}

\textcolor{black}{Last but not least, we note that the differences in the Hubbard and exchange parameters in PNO and CCO can be understood as arising from the following facts: (i) differences in the Hubbard interaction $U$ (monopole interaction) across transition-metal ions in crystals are primarily governed by the nuclear charge and are only weakly affected by covalency effects; (ii) differences in the Hund's exchange $J_H$ (multipole interactions) between different transition-metal ions are predominantly controlled by the degree of covalency via the nephelauxetic effect~\cite{schaffer1958nephelauxetic}; (iii) the Hubbard $U$ is larger in copper than in nickel because of the higher nuclear charge of copper; and (iv) the Hund's exchange $J_H$ is reduced in CCO relative to PNO due to the substantially stronger covalency in CCO~\cite{botana2020similarities}.}

\bibliography{BiblioFR}

\section{Funding information}
 F.R., G.M., L.M., M.Z., M.S. and G.G. acknowledge support by the projects PRIN2017 Quantum-2D–ID 2017Z8TS5Band PRIN 2020 QT-FLUO ID 20207ZXT4Z of the Ministry for University and Research (MIUR) of Italy. D.P. acknowledges the ANR–FRANCE (French National Research Agency) for its financial support of the T-ERC project ORBIFUN ANR-23-ERCS-0003-01. This work was partially performed within the MUSA‐Multilayered Urban Sustainability Action project, funded by the European Union‐NextGenerationEU, under the National Recovery and Resilience Plan (NRRP) Mission 4 Component 2 Investment Line 1.5: Strengthening of research structures and creation of R\&D,“innovation ecosystems” set up of “territorial leaders” in R\&D. 
 \textcolor{black}{K.W. acknowledges the support
of National Science Centre in Poland under Project No.
2024/55/B/ST3/03144 and No. 2021/43/B/ST3/02166.} The work by G.M. was jointly supported by Politecnico di Milano and European X-ray Free Electron Laser Facility GmbH.

\section{Acknowledgments}

We thank Marco Moretti Sala \textcolor{black}{and Mark Dean} for valuable discussion. This research used ESRF beamline ID32 under the Proposal No. HC5438, using the ERIXS spectrometer designed jointly by the ESRF and Politecnico di Milano.

\section{Author contributions}

F.R., G.M., L.M.,  N.B., M.S. and G.G. conceived and performed the RIXS measurements, with suggestions from R.A. and D.P.
H.S. and D.P. grew and characterized the PNO thin films; D.D.C. grew and characterized the CCO films.
F.R. analyzed the RIXS experimental data and performed the fittings.
F.R., M.S. and G.G. discussed and interpreted the results.
\textcolor{black}{K.W. developed the theoretical model for orbiton propagation.}
M.Z. performed single-ion calculations.
F.R. and G.G. wrote the manuscript with major suggestions from R.A. and K.W., and contributions from all authors.

\section{Competing interests}
The authors declare no competing interests.

\end{document}